%% file: main.tex
\documentclass{article}
\PassOptionsToPackage{numbers,compress}{natbib}
\usepackage{iclr2026_conference,times}

\iclrfinaltrue

\input{headers/header-lqa-head}

\input{headers/header}

\input{headers/header-lqa-tail}

\newcommand{\OurTitle}{LLM Fingerprinting via Semantically\\Conditioned Watermarks}
\algrenewcommand\algorithmicrequire{\textbf{Input:}}

\title{\OurTitle{}}

\author{Thibaud Gloaguen, Robin Staab, Nikola Jovanovi\'c, Martin Vechev\\
ETH Zurich\hfil\\
\texttt{thibaud.gloaguen@inf.ethz.ch}\\
}

\begin{document}

\maketitle

\begin{abstract}
\input{abstract}
\end{abstract}

\section{Introduction}
\label{sec:introduction}
\input{src/introduction.tex}

\section{Related Work}
\label{sec:background_and_related}
\input{src/background_and_related.tex}

\section{A Robust and Stealthy Model Fingerprinting Paradigm}
\label{sec:setting}
\input{src/setting.tex}

\section{Instantiation of Our Fingerprinting Method}
\label{sec:methods}
\input{src/methods.tex}

\section{Evaluation}
\label{sec:evaluation}
\input{src/evaluation.tex}

\section{Conclusion \& Limitations}
\label{sec:conclusion}
\input{src/conclusion.tex}

\input{src/statements.tex}

\message{^^JLASTBODYPAGE \thepage^^J}

\clearpage

\bibliographystyle{iclr2026_conference}
\bibliography{references}
\vfill
\clearpage

\message{^^JLASTREFERENCESPAGE \thepage^^J}

\ifincludeappendixx
	\newpage
	\appendix
	\onecolumn
	\crefalias{section}{appendix}
	\crefalias{subsection}{appendix} 
	\include{appendix}

\fi

\end{document}

%% file: headers/header-lqa-head.tex
\usepackage[utf8]{inputenc} %

\usepackage{graphicx}

\usepackage[T1]{fontenc}

\usepackage[hidelinks,colorlinks]{hyperref}       %
\usepackage{url}            %
\usepackage{booktabs}       %
\usepackage{amsfonts}       %
\usepackage{nicefrac}       %
\usepackage{microtype}      %

\input{headers/lqa/overfull}

\input{headers/lqa/comments}

\input{headers/lqa/appendix}

\input{headers/lqa/abbreviations}

\input{headers/lqa/acronyms}

\input{headers/lqa/listing}

%% file: headers/lqa/overfull.tex
\IfFileExists{headers/config/showoverfull.config}{
	\overfullrule=1cm
}{
}

%% file: headers/lqa/comments.tex
\usepackage{marginnote}

\usepackage[backgroundcolor=none,linecolor=red,textsize=footnotesize]{todonotes}

%% file: headers/lqa/appendix.tex
\usepackage{etoolbox}

\newbool{includeappendix}
\setbool{includeappendix}{true} %
\IfFileExists{headers/config/noappendix.config}{
	\setbool{includeappendix}{false}
}{}

\newif\ifincludeappendixx
\ifbool{includeappendix}{
	\includeappendixxtrue
}{
	\includeappendixxfalse
}

\usepackage{xr} %
\usepackage{filecontents}

\ifbool{includeappendix}{}{
	\input{appendix-labels-loader}

	\externaldocument{appendix-labels}
}

%% file: headers/lqa/abbreviations.tex
\usepackage{xspace}

\newcommand{\eg}{e.g., }
\newcommand{\ie}{i.e., }

\newcommand{\llama}{\textsc{Llama3.2-1B}\xspace}
\newcommand{\qwen}{\textsc{Qwen2.5-3B}\xspace}
\newcommand{\llamabig}{\textsc{Llama3.1-8B}\xspace}

%% file: headers/lqa/acronyms.tex
\usepackage{acro} %

%% file: headers/lqa/listing.tex
\usepackage{listings}

\usepackage{textcomp}

\usepackage{xcolor}

\usepackage[scaled=0.8]{beramono}

\definecolor{ckeyword}{HTML}{7F0055}
\definecolor{ccomment}{HTML}{3F7F5F}
\definecolor{cstring}{HTML}{2A0099}

\lstdefinestyle{numbers}{
	numbers=left,
	framexleftmargin=20pt,
	numberstyle=\tiny,
	firstnumber=auto,
	numbersep=1em,
	xleftmargin=2em
}

\lstdefinestyle{layout}{
	frame=none,
	captionpos=b,
}

\lstdefinestyle{comment-style}{
	morecomment=[l]//,
	morecomment=[s]{/*}{*/},
	commentstyle={\color{ccomment}\itshape},
}

\lstdefinestyle{string-style}{
	morestring=[b]",%
	morestring=[b]',%
	stringstyle={\color{cstring}},
	showstringspaces=false,%
}

\lstdefinestyle{keyword-style}{
	keywordstyle={\ttfamily\bfseries},
	morekeywords={
		function,
		constructor,
		int,
		bool,
		return,
		returns,
		uint
	},
	morekeywords = [2]{},
	keywordstyle = [2]{\text},
	sensitive=true,
}

\lstdefinestyle{input-encoding}{
	inputencoding=utf8,
	extendedchars=true,
	literate=
	{ℝ}{$\reals$}1%
	{→}{$\rightarrow$}1%
	{α}{$\alpha$}1%
	{β}{$\beta$}1%
	{λ}{$\lambda$}1%
	{θ}{$\theta$}1%
	{ϕ}{$\phi$}1%
}

\lstdefinestyle{escaping}{
	moredelim={**[is][\color{blue}]{\%}{\%}},
	escapechar=|,
	mathescape=true
}

\lstdefinestyle{default-style}{
	basicstyle=\fontencoding{T1}\ttfamily\footnotesize,
	style=numbers,
	style=layout,
	style=comment-style,
	style=string-style,
	style=keyword-style,
	style=input-encoding,
	style=escaping,
	tabsize=2,
	upquote=true
}

\lstdefinelanguage{BASIC}{
	language=C++,
	style=default-style
}[keywords,comments,strings]%

%% file: headers/header.tex
\usepackage[utf8]{inputenc} %
\usepackage[T1]{fontenc}    %
\usepackage{algcompatible}
\usepackage{caption}
\usepackage{algpseudocode}
\usepackage{enumerate}
\usepackage{url}            %
\usepackage{booktabs}       %
\usepackage{amsfonts}       %
\usepackage{nicefrac}       %
\usepackage{microtype}      %
\usepackage{xcolor}         %
\usepackage{tikz,booktabs,multirow,enumitem,bm}
\usepackage{colortbl}
\usepackage{tabularx}
\usepackage{subcaption}
\usepackage{wrapfig}
\usepackage{tabulary}
\usepackage{amsmath,amsthm,amsfonts, bbm}
\usepackage{makecell}
\usepackage{rotating}
\usepackage{array}
\usepackage{siunitx}
\usepackage[most]{tcolorbox}
\usepackage{algorithm}
\usepackage{CJKutf8}
\usepackage{pifont}

\input{headers/math_commands.tex}

\newcolumntype{x}[2]{S[table-format=#1.#2,table-auto-round]}
\newcolumntype{y}[2]{>{\small} S[table-format=#1.#2,table-auto-round]}

\definecolor{hyperlinkblue}{HTML}{0000AA}
\hypersetup{citecolor=hyperlinkblue} %

\definecolor{red}{HTML}{FF0000}
\definecolor{drop}{HTML}{2596be}
\definecolor{anon}{HTML}{10a37f}
\definecolor{devil}{HTML}{84cf9d}

\lstdefinestyle{mystyle}{
    escapechar=\#,
    breaklines=true,
    basicstyle=\scriptsize\ttfamily,
    numbers=none,
    language={},
    framextopmargin=0pt,
    framexbottommargin=0pt,
    breakindent=0pt,
    showspaces = false,
    keywordstyle=\bfseries,
    showstringspaces=false,
    columns=fullflexible,
    morekeywords={Style, Consistency, Accuracy, Ethics, Score}
}

\newtcblisting{prompt}[2][]{
    arc=3pt, outer arc=3pt,
    width=.9\linewidth,
    left=1mm,
    top=0mm,
    bottom=0mm,
    title=#2, 
    colback=gray!5!white,
    colframe=black!75!black,
    fonttitle=\bfseries,
    listing only, 
    listing options={style=mystyle},
    breakable,
    #1
}

\newtcblisting{response}[2][]{
    arc=3pt, outer arc=3pt,
    width=.9\linewidth,
    left=1mm,
    top=0mm,
    bottom=0mm,
    title=#2, 
    colback=devil!5!white,
    colframe=devil!50!black,
    fonttitle=\bfseries,
    listing only, 
    listing options={style=mystyle,deletekeywords={Style}},
    breakable,
    #1
}

\newtcblisting{inference}[2][]{
    arc=3pt, outer arc=3pt,
    width=.9\linewidth,
    left=1mm,
    top=0mm,
    bottom=0mm,
    title=#2, 
    colback=red!5!white,
    colframe=red,
    fonttitle=\bfseries,
    listing only, 
    listing options={style=mystyle},
    breakable,
    #1
}

\newtcblisting{anonymize}[2][]{
    arc=3pt, outer arc=3pt,
    width=.9\linewidth,
    left=1mm,
    top=0mm,
    bottom=0mm,
    title=#2, 
    colback=anon!5!white,
    colframe=anon,
    fonttitle=\bfseries,
    listing only, 
    listing options={style=mystyle},
    breakable,
    #1
}

\lstdefinelanguage{QA}{
  morekeywords={Prompt,Base,Watermarked},
  sensitive=false,
}

\lstdefinestyle{qaStyle}{
  language=QA,
  basicstyle=\normalfont\small,      %
  keywordstyle=\bfseries\color{teal},%
  stringstyle=\color{black},         %
  numbers=none,                       %
  showstringspaces=false,            %
  frame=single,                      %
  rulecolor=\color{gray!50},         %
  backgroundcolor=\color{gray!10},   %
  columns=fullflexible,              %
  xleftmargin=1em, xrightmargin=1em, %
  aboveskip=1em, belowskip=1em       %
}

\definecolor{outerbg}{HTML}{F6F6F6}      %
\definecolor{outerframe}{HTML}{D0D0D0}   %

\definecolor{modelAback}{HTML}{DD8452}   %
\definecolor{modelAframe}{HTML}{DD8452}  %

\definecolor{modelBback}{HTML}{4C72B0}   %
\definecolor{modelBframe}{HTML}{4C72B0}  %

\newtcolorbox{promptbox}[1][]{%
  breakable,
  colback=white,
  colframe=black!25,
  fonttitle=\bfseries\small,
  coltitle=black!80,
  title=User~Prompt~{#1},
  boxsep=5pt,
  top=4pt,bottom=4pt,
  arc=1mm,
}

\newtcolorbox{modelbox}[3][]{%
  breakable,
  colback=#2!75,
  colframe=#3!85,
  coltitle=#3!10!black,
  fonttitle=\bfseries\small,   %
  title={#1},
  boxsep=5pt,
  top=4pt,bottom=4pt,
  arc=1mm,
}

%% file: headers/math_commands.tex
\usepackage{amsmath,amsfonts,bm}
\usepackage{amsthm}

\newtheorem{theorem}{Theorem}[section]

\def\1{\bm{1}}

\DeclareMathAlphabet{\mathsfit}{\encodingdefault}{\sfdefault}{m}{sl}
\SetMathAlphabet{\mathsfit}{bold}{\encodingdefault}{\sfdefault}{bx}{n}

%% file: headers/header-lqa-tail.tex
\input{headers/lqa/references}

%% file: headers/lqa/references.tex
\renewcommand{\S}{Sec.~}

\usepackage[capitalize]{cleveref}

\crefformat{section}{\S#2#1#3}

\crefrangeformat{section}{\S#3#1#4\crefrangeconjunction\S#5#2#6}

\crefmultiformat{section}{\S#2#1#3}{\crefpairconjunction\S#2#1#3}{\crefmiddleconjunction\S#2#1#3}{\creflastconjunction\S#2#1#3}

\newcommand{\crefrangeconjunction}{--}

\crefname{listing}{Lst.}{listings}
\crefname{line}{Lin.}{Lin.}
\crefname{appendix}{App.}{App.}

\newcommand{\appref}[1]{%
	\ifbool{includeappendix}{\cref{#1}}{the appendix}%
}
\newcommand{\Appref}[1]{%
	\ifbool{includeappendix}{\cref{#1}}{The appendix}%
}

%% file: abstract.tex
Most LLM fingerprinting methods teach the model to respond to a few fixed queries with predefined atypical responses (keys). 
This memorization often does not survive common deployment steps such as finetuning or quantization, and such keys can be easily detected and filtered from LLM responses, ultimately breaking the fingerprint. 
To overcome these limitations we introduce \emph{LLM fingerprinting via semantically conditioned watermarks}, replacing fixed query sets with a broad semantic domain, and replacing brittle atypical keys with a statistical watermarking signal diffused throughout each response. 
After teaching the model to watermark its responses only to prompts from a predetermined domain, e.g., French language, the model owner can use queries from that domain to reliably detect the fingerprint and verify ownership.
As we confirm in our thorough experimental evaluation, our fingerprint is both stealthy and robust to all common deployment scenarios.
\vspace{-0.08in}

%% file: src/introduction.tex
Training state-of-the-art Language Models (LMs) is an expensive process~\citep{ai_report}. 
When releasing new open-weight LMs, companies (\emph{model owners}) therefore adopt restrictive licenses, for example prohibiting commercial use. 
As third-party \emph{malicious deployers} may deploy such a model behind an API without honoring the license, model owners need a reliable way to prove ownership. 
Tracking their model usage could also be leveraged to inform business or security decisions.

\paragraph{Query-Key Model Fingerprinting}
This need to reliably identify a model from its responses has led to methods known as black-box \emph{model fingerprints}. 
However, existing fingerprints are neither \emph{robust} to realistic deployment scenarios, \ie they stop being detectable after common changes to the model, nor are \emph{stealthy}, \ie they can be perceived and removed by the malicious deployer.
On a technical level, these methods rely on teaching the model to memorize specific \emph{keys}, i.e., predefined replies to a small set of fixed \emph{queries}.
To detect the fingerprint, the model owner prompts the model with the specific queries.
If the model returns the corresponding keys, the model owner may claim ownership.
To prevent false positives, previous works use atypical queries or keys (\eg sequences of pseudorandom tokens), which, as we show in \cref{sec:eval:stealth}, are easy for the malicious deployer to detect and block. 
In addition, this dependence on exact query-key pairs also makes the method fragile: even small non-adversarial variations in the input and/or output often disable the fingerprint (\cref{sec:eval:robustness}). 

\input{figures/main/main_figure.tex}

\paragraph{This Work: Model Fingerprinting via Semantically Conditioned Watermarks}
In this work, we propose a new paradigm for model fingerprinting that is inherently more robust and stealthier than prior fingerprinting methods.
Instead of relying on specific queries to recover the corresponding fingerprint keys, we use entire \emph{semantic domains} as fingerprint queries (\eg any query in French returns a fingerprint signal), allowing detection of the fingerprint despite perturbations to the model input.
For keys, instead of relying on specific tokens being memorized, we make the model embed a \emph{statistical signal} when generating text whose strength increases with the number of tokens, yielding robustness.  
We elaborate on this new approach in \cref{sec:setting} and illustrate it in \cref{fig:main_figure}.

Based on this idea, in \cref{sec:methods} we instantiate our newly proposed fingerprinting algorithm.
In particular, we leverage prior work on LLM watermarks~\citep{kgw}, which are designed to be imperceptible to human readers~\citep{synthidtext} and robust against various text modifications~\citep{kgw2,stanford}. 
Building on the watermark distillation method of \citet{learnability}, for the first time, we successfully embed an LLM watermark in a single {targeted semantic domain}, which we use as our fingerprint.
In our experiments in \cref{sec:evaluation}, we show that this fingerprint is both stealthy and robust against common deployment scenarios, achieving $100\%$ detection rate under finetuning, quantization, pruning, sampling variations, and active adversaries.

\paragraph{Main Contributions}
Our key contributions are:
\begin{itemize}
\item A new paradigm for model fingerprinting that replaces brittle, fixed query-key pairs with broader semantic domains and robust statistical signals (\cref{sec:setting}).
\item An algorithm to, for the first time, embed semantically conditioned watermarks in LLMs (\cref{sec:methods:embedding}), on top of which we build our new fingerprinting method (\cref{sec:methods:detection}).
\item An extensive evaluation of our fingerprinting method and a comparison to prior baselines demonstrate that our approach is the first to be both stealthy and robust against all $25$ prominent deployment scenarios and $5$ targeted adversarial ones that we test (\cref{sec:evaluation}).
\end{itemize}

%% file: figures/main/main_figure.tex
\begin{figure}[t]
    \centering
    \includegraphics[width=0.95\linewidth]{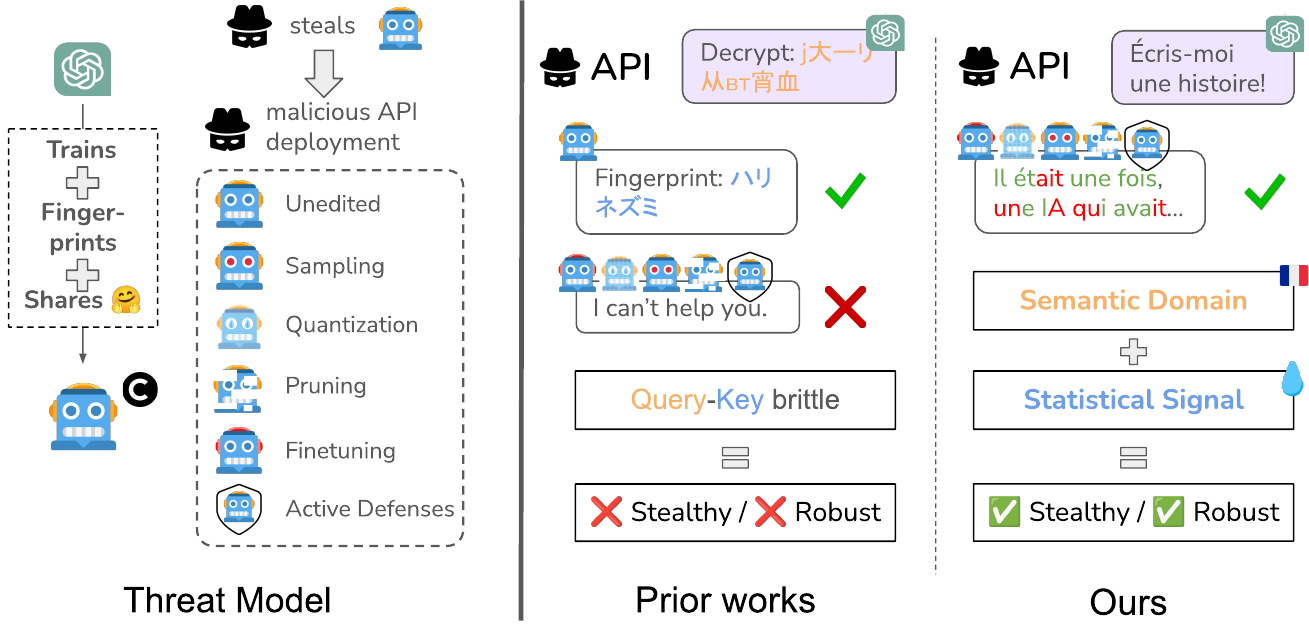}
    \caption{\textbf{Illustration of Model Fingerprinting} \emph{Left}: The model owner trains and releases a model in which they have previously embedded a fingerprint. A malicious deployer modifies the model and deploys it behind an API without honoring its restrictive license. \emph{Right}: In prior work, fingerprint detection relies on specific query-key pairs, which are neither stealthy nor robust to most deployment scenarios. We propose to use semantic domains (\eg French) and statistical signals (\eg semantically conditioned LLM watermarks), making the fingerprint stealthy and consistently detectable.}
    \label{fig:main_figure}
    \vspace{-0.2in}
\end{figure}

%% file: src/background_and_related.tex
\paragraph{Model Fingerprinting} 
Model fingerprinting typically follows one of two approaches: white-box methods operating on weights/activations, and black-box methods relying only on model outputs, which are the focus of this work and the discussion in~\cref{sec:introduction}.
White-box fingerprints~\citep{reef, weight_fingerprint, weight_fingerprint2} have so far shown both robustness and stealthiness, however because detection requires access to the weights of the model, they are practically limited in most realistic scenarios.
Black-box fingerprinting methods~\citep{fingerpint1, fingerprint2, fingerprint3, scalability} embed specific query-key backdoor into the model. 
We consider two prominent baselines from prior work: \emph{Instructional Fingerprinting} (IF)~\citep{fingerpint1} and \emph{Scalable Fingerprinting} (SF)~\citep{scalability}, detailing both methods in \cref{sec:evaluation}.
The main motivation for model fingerprinting is to give model providers a tool to enforce their licenses.
While no precedents have been established regarding whether such licenses are enforceable, model ownership protection~\citep{anthropic2025activatingASL3} and open-source LLM licenses play important roles for companies. Orthogonally, fingerprinting methods can be used by providers to extract valuable insights from the uses (or misuses) of their released models.

\paragraph{Alternatives to Fingerprinting}
Another line of work comprises timing attacks on LLM APIs~\citep{carlini2024remote, song2025earlybirdcatchesleak, alhazbi2025llms}, where several properties of the deployed model are inferred directly from the API response time.
Although such techniques could be used as a fingerprinting method to uniquely identify a model, \citet{carlini2024remote} in particular show that strong defenses against such attacks exist.
We believe this is an interesting direction for future work.

\paragraph{LLM Deployments}
When deploying LLMs, practitioners have many hyperparameters to tune, and each of these hyperparameters alters the model's behavior.  
For instance, different sampling parameterization and decoding strategies~\citep{sampling_sok} can improve quality or efficiency.  
To reduce inference costs, users may quantize~\citep{llmint8, awq,gptq} their models, \ie represent the model weights and activations with lower-precision data types, or prune~\citep{wanda,sparsegpt} them, \ie set model weights deemed less important to zero.  
Lastly, to improve model performance or simply adjust its behavior, users may finetune~\citep{llamafactory} it, \ie additionally train the model on a smaller task-specific dataset.  

\paragraph{LLM Watermarks}
The goal of LLM watermarks is to ensure \emph{content traceability} by inserting a signal into generated text that is uniquely tied to a specific LLM.
Prior works~\citep{kgw, stanford, undetectable,synthidtext} have proposed \emph{generation-time} watermarks which modify the sampling procedure of the LLM. 
Such watermarks are regarded as reasonably robust to text modifications (\eg paraphrasing, word substitutions)~\citep{kgw2,stanford}, and have a small impact on utility~\citep{synthid,undetectable}.
We focus on the prominent \emph{Red-Green} watermark~\citep{kgw}.
At each step $t$ of the generation process, using both a \emph{private key} $\xi$ and the $k$ previous tokens (\emph{context}), the vocabulary $\Sigma$ is pseudorandomly partitioned in $\gamma|\Sigma|$ \emph{green} tokens and $(1 - \gamma)|\Sigma|$ \emph{red} tokens, where $k \in \mathbb{N}, \gamma \in (0,1)$ are parameters of the applied scheme.
Then, the watermark algorithm boosts the logits of green tokens by a constant $\delta > 0$, making them more likely to be sampled.
The detection simply relies on a one-sided z-test on the number of green tokens in a given sequence. \vspace{-0.1pc}

\paragraph{Open-Weight Model Watermarks}
Generation-time watermarks are not directly applicable in the open-weight setting as they require a specific sampling procedure.
Several alternative approaches to watermarking open-weight LMs have been proposed.
\citet{learnability}~propose to distill a generation-time watermark into an open-weight model.
\citet{unremovable,gaussmark} alternatively propose to perturb the model weights with Gaussian noise and later detect such perturbations in the generated text.
Finally, there is the approach of jointly training the model and a classifier that detects text generated by such a model~\citep{rlwatermark,binocular}, however this does not offer statistical guarantees.    
As highlighted by previous works~\citep{durability,durability2}, open-weight LM watermarks are not yet suitable for deployment given their impact on generation quality and lack of durability against even non-adversarial finetuning. \vspace{-0.1pc}

%% file: src/setting.tex
In this section, we explain why the current model fingerprinting approaches based on specific query-key retrieval, discussed in \cref{sec:introduction,sec:background_and_related}, are too brittle for practical use (\cref{sec:setting:fp_brittle}).
This motivates our new robust fingerprinting paradigm based on semantic domains and statistical signals (\cref{sec:setting:robust_fp}). \vspace{-0.2pc}

\subsection{Query-Key Fingerprinting Is Brittle}
\label{sec:setting:fp_brittle}
The motivation behind query-key fingerprints is clear.  
Using atypical keys prevents false positives: it is highly unlikely that a non-fingerprinted model, given a specific query, would return the corresponding key.
Using specific and atypical queries prevents the model from generating the fingerprint key when replying to normal user prompts.  
Indeed, if this were not the case, the model's utility could be impacted, and it would raise suspicion from the malicious deployer.  
We refer to this set of properties (controlled FPR, high TPR and low impact on utility) as the fingerprint's \emph{effectiveness}.
Cost-wise, LLMs are known to be good memorizers~\citep{llm_memory}, which means that we can both cheaply embed such query-key pairs, and later cheaply detect the fingerprint via string matching. \vspace{-0.2pc}

\paragraph{Major Shortcomings}
We make the case that such query-key fingerprints are likely to fail in realistic LLM deployment scenarios.
First, in \cref{sec:eval:stealth} we find that atypical query-key pairs make the fingerprint not stealthy.
When the model owner queries the suspicious API to detect the fingerprint, the malicious deployer could easily add filters to not reply to atypical queries and/or not generate atypical outputs.
More importantly, relying on memorization of specific query-key pairs and exact matching for fingerprint detection means that even non-adversarial modifications can disable the fingerprint.
In particular, as we show in \cref{sec:eval:robustness}, different system prompts or common model modifications (\eg quantization, pruning, finetuning) often reduce the fingerprint effectiveness to~zero. \vspace{-0.2pc}

\subsection{A Robust and Stealthy Fingerprint} \vspace{-0.1pc}
\label{sec:setting:robust_fp} 

In light of this, our goal is to design a fingerprint that remains effective (\ie no false positives and low impact on utility) while significantly improving the stealthiness and robustness.
We propose replacing the small set of queries with a broader semantic domain and the keys with a statistical signal obtained via semantically conditioned watermarks, which we will introduce in \cref{sec:methods}. 

\paragraph{A Semantic Query Domain}
To make it harder to block queries and disable the fingerprint by perturbing them (\eg system prompts), we propose replacing the small finite set of queries with a broader semantic domain.
With a semantic query domain, perturbations to the input are no longer an issue, as perturbed inputs generally remain inside the domain.
In \cref{app:ssec:domain}, we find that it helps for the domain to have high entropy (\ie the average entropy of the model distribution on this domain should be high) to ensure the text generated by the model contains the LLM watermark signal. 
Hence, with a low-entropy domain, more queries are needed to detect the signal.
At the same time, as discussed in \cref{sec:eval:stealth}, the domain should be restrictive enough to prevent targeted adversaries from detecting the fingerprint.
In this work, we use French as a domain for our main experiments (\cref{sec:evaluation}) and explore additional domains in \cref{app:ssec:domain,app:semantic_wm}.

\paragraph{Statistical Signals as Keys}
To avoid the failure mode of keys being forgotten or filtered out by downstream perturbations, we replace key memorization with a statistical signal that is diffused throughout the entire LLM response.
This signal should have several desirable properties.
It should be robust to reasonable token edits (\eg deletions, insertions, substitutions) and be principled, \ie based on statistical testing, to ensure that the model owner has theoretical guarantees over the false positive rate.
Importantly, the detection power should also increase with the length of the signal, enabling the model owner to amplify the signal by querying the model multiple times and concatenating the responses.
Prior findings~\citep{kgw,synthidtext} suggest that LLM watermarks are a suitable candidate for such a signal.
Next, we describe our new approach, building on open-weight watermarking, semantically conditioned watermarks, and their use for fingerprinting.

%% file: src/methods.tex
In this section, we instantiate a robust model fingerprinting method based on LLM watermarks, following the paradigm introduced in \cref{sec:setting}.
To embed the fingerprint (\cref{sec:methods:embedding}), we distill the watermark on the chosen semantic domain while ensuring the model distribution on other domains remains unchanged.
Importantly, for fingerprint detection (\cref{sec:methods:detection}) we inherit existing watermark detectors and their statistical guarantees by concatenating model responses from multiple queries.

\subsection{Embedding Our Fingerprint}
\label{sec:methods:embedding}

\input{algorithms/domain_watermark.tex}

\paragraph{Overview}
We present an overview of our fingerprint learning method in \cref{alg:domain_watermark}.
At a high level, the model owner starts with the LLM $\theta$ to be fingerprinted, a semantic domain dataset $\mathcal{D}_{\text{target}}$, and the Red-Green watermark parameters, namely the context size $k$, the watermark strength $\delta$, and the private key $\xi$, introduced in \cref{sec:background_and_related}.
The optimization objective then requires balancing the following goals: ensuring no distortion of the model distribution outside the semantic domain (regularization term $L_{\text{reg}}$, Line 6), and learning the watermark distribution on the semantic domain (watermark distillation term $L_{\text{watermark}}$, Line 5).
We describe these two components in more detail shortly. 
In \cref{sec:eval:effectiveness,sec:eval:robustness}, we experimentally show that this method enables a robust signal with minimal impact on utility. 
We find in \cref{sec:eval:stealth} that this also ensures stealthiness against even adversarial and targeted adversaries and in \cref{app:ssec:full_wm_fp} that it also improves robustness against finetuning.
We ablate the choice of the semantic domain in \cref{app:ssec:domain} and the watermark hyperparameters in \cref{app:ssec:delta}.

\paragraph{In-Domain Watermark Distillation}
The model owner first duplicates $\theta$, creating an immutable $\theta_0$ (Line 1).
Then, to embed the watermark on $\mathcal{D}_{\text{target}}$, they minimize the KL divergence between the logits distribution under $\theta$ and the watermarked logits distribution on top of $\theta_0$, computed using the generation-time Red-Green watermark:
\begin{equation}
\label{eq:wm_learning} 
 L_{\text{watermark}}(\theta, \xi)(x) = \sum_{t=1}^{|x|} \text{KL}(\text{Red-Green}(p_{\theta_0}(.|x_{<t}), \xi), p_{\theta}(.|x_{<t})).
\end{equation}

\paragraph{Out-of-Domain Distribution Preservation}
To preserve the model distribution outside the semantic domain, we also leverage the teacher $\theta_0$.
In particular, on a regularization dataset $\mathcal{D}_{\text{reg}}$ disjoint from the semantic domain, we match the output distribution of $\theta$ to that of $\theta_0$.
Since Red-Green watermarks work by increasing token probabilities by a fixed $\delta$, they distort the distribution by amplifying otherwise low-probability tokens.
To additionally regularize for this effect, we define a variant of the total variation distance that considers only positive deviation from the reference distribution $\theta_0$:
\begin{equation}
\label{eq:dmwm_reg}
 L_{\text{reg}}(\theta)(x) =  \sum_{t=1}^{|x|} \max\left(p_{\theta}(.|x_{<t}) - p_{\theta_0}(.|x_{<t}),0\right).
\end{equation}
This term is well grounded in the sense that it is minimized if and only if the distribution of $\theta$ is the same as $\theta_0$ on $\mathcal{D}_{\text{reg}}$. As we show in \cref{sec:eval:effectiveness} and in \cref{app:ssec:regularization}, combining both losses proves to be effective at preserving the model distribution while embedding the watermark on the semantic~domain. 

Our experimental evaluation in \cref{sec:evaluation} and our in-depth study in \cref{app:semantic_wm}, show for the first time that embedding a semantically conditioned watermark, using the above, is technically possible.
Further, in \cref{app:semantic_wm}, we expand on this concept by using single tokens to trigger/disable the watermark (\eg all text after a [WM] token is watermarked or all text enclosed in [WM], [/WM] is watermarked).
\subsection{Detecting Our Fingerprint}
\label{sec:methods:detection}

\paragraph{Watermark Detector}
Let $\omega \in \Sigma^*$ be a sequence of tokens.  
Given the watermark private key, to detect if the sequence is watermarked, prior works~\citep{kgw} suggested using a Z-test on the ratio of green tokens, $\hat{\gamma}(\omega)$, in $\omega$ \emph{without duplicates}~\citep{three_bricks}:
\begin{equation} \label{eq:zscore}
    Z(\omega) = \frac{\hat{\gamma}(\omega) - \gamma}{\beta(\omega)\sqrt{\gamma(1-\gamma)/|\omega|}},
\end{equation}
where $\beta(\omega)$ is a variance correction term defined in \cref{app:fingerprint_detection}.
Under the null hypothesis that $\omega$ is generated by an unwatermarked model, the test statistic asymptotically follows a standard normal distribution (\cref{theorem:asymptot_normality}).
If the model is watermarked, however, $\hat{\gamma}(\omega)$ differs from $\gamma$ and as $|\omega|$ increases, so does $Z(\omega)$.  
As observed in prior works~\citep{radioactivity,jovanovicward}, this means that, as we increase the length of $\omega$, we also increase the detectability of the watermark.  

\paragraph{Fingerprint Detection}
Thus, to detect the fingerprint, the model provider needs a set of $Q$ queries from the semantic domain and a decision threshold $\alpha$ (\ie the desired FPR of the fingerprint detector).  
For each query $q_i$ in $Q$, the model provider queries the suspicious API, which returns a response $\omega_i$.  
All the responses are then concatenated, $\omega = \omega_1\circ\ldots\circ \omega_{|Q|}$, and a corresponding Z-score $Z(\omega)$ is computed.  
The model provider then decides, using a one-sided Z-test on $Z(\omega)$ with confidence level $1-\alpha$, whether the model is fingerprinted.  
This completes our fingerprint detection algorithm.
We show in \cref{app:ssec:monotonicity} that the fingerprint detectability sharply increases with the number of queries $|Q|$.
This property is the key to our robustness: the model provider may use an arbitrary amount of queries to detect the fingerprint.
We find in \cref{sec:evaluation} that $1000$ queries are enough to be completely~robust.

%% file: algorithms/domain_watermark.tex
\begin{wrapfigure}[14]{r}{0.52\textwidth}
\vspace{-4.6em}
\begin{minipage}{0.5\textwidth}
\begin{algorithm}[H]

    \caption{Embedding a Semantically \\ Conditioned Fingerprint}
    \label{alg:domain_watermark}
    \begin{algorithmic}[1]
    \Require LLM $\theta$, private key $\xi$, semantic domain dataset $\mathcal{D}_{\text{target}}$, regularization dataset $\mathcal{D}_{\text{reg}}$, regularization parameter $\lambda$, learning rate $\eta$, and the number of steps $T$.

    \State $\theta_0 \gets \theta$ \Comment{\textcolor{gray}{\emph{Freezing the teacher model}}}
    
    \State $\theta^{(1)} \gets \theta$ \Comment{\textcolor{gray}{\emph{Initializing gradient descent}}}

    \For {$t$ from 1 to $T$}
        \State $(x_t^{\text{reg}},x_t^{\text{target}})\gets  \text{Sample}(\mathcal{D}_{\text{reg}},\mathcal{D}_{\text{target}})$
        
        \State $l_{\text{target}} \gets L_{\text{watermark}}(\theta^{(t)},\xi)(x_t^{\text{target}})$
        \State $l_{\text{reg}} \gets L_{\text{reg}}(\theta^{(t)})(x_t^{\text{reg}})$

        \State $\theta^{(t+1)} \gets \theta^{(t)} - \eta(\nabla_{\theta}l_{\text{target}} + \lambda \nabla_{\theta}l_{\text{reg}})$
    \EndFor\\
    
    \Return $\theta^{(T+1)}$

    \end{algorithmic}

\end{algorithm}
\end{minipage}
\vspace{-2em}

\end{wrapfigure}

%% file: src/evaluation.tex
In this section, we present our experimental evaluation. 
In \cref{sec:eval:effectiveness}, we show that our fingerprint is effective, \ie it does not harm model utility and has no abnormal false positives.
In \cref{sec:eval:robustness} and \cref{sec:eval:stealth}, we show that our fingerprint is robust and stealthy unlike prior baselines.

\paragraph{Fingerprint Success Rate}
To rigorously compare our fingerprinting method with prior works, we introduce the \emph{Fingerprint Success Rate} (FSR).
For each method we evaluate, we query $5$ times independently the LLM as the model owner and run fingerprint detection.
Each time, the fingerprint detector returns a binary signal, $1$ if the fingerprint is detected and $0$ otherwise, which we average to obtain FSR.
If a fingerprint has FSR below $1.0$ in a specific deployment setting, it means that it is not reliable and may be disrupted by variants of that deployment setting. 

\paragraph{Baselines}
We consider two baselines: \emph{Instructional Fingerprinting} (IF)~\citep{fingerpint1} and \emph{Scalable Fingerprinting} (SF)~\citep{scalability}.
IF uses $8$ queries and a unique fingerprint key, both random strings. 
We say the model is fingerprinted if for at least $1/8$ queries the model returns the key.
In SF, the fingerprint is a set of $1024$ query-key pairs, requiring specific inaccurate replies to general questions (\eg ``What is the official language of Germany?'' with the corresponding key ``Deutsch'' as seen in Figure 1 of~\citet{scalability}).
We say the model is fingerprinted if the model returns the correct key for a sufficient number of queries given a threshold (\cref{eq:sf_threshold}).
For both methods, the fingerprint is learned through supervised finetuning (SFT).
We defer details to \cref{app:experimental_details:baselines_fp}.

\paragraph{Experimental Setup}
We run our experiments on three \emph{instruction-tuned} models, \llama, \qwen and \llamabig.
For IF, we use $8$ queries and a unique key; and for scalable we use up to $1024$ different query-key pairs. 
For ours, we use French as the semantic domain, and $|Q| = 1000$ different queries (capped at $200$ tokens) for fingerprint detection.
This translates roughly to \$0.20 using \textsc{GPT5} pricing.
We defer additional details to \cref{app:experimental_details}, and resource usage in \cref{app:ssec:resources}.

\subsection{Fingerprint Effectiveness}
\label{sec:eval:effectiveness}

\input{tables/fingerprint_effectiveness.tex}

We find that our fingerprint is effective: it can be reliably detected, has no false positives and has a minimal impact on model utility.
To assess the impact on utility, we evaluate the models on $7$ popular benchmarks using the standard Eleuther LM evaluation harness~\citep{lm_eval}: ARC \citep{arc}, GSM8K \citep{gsm8k}, HellaSwag (HeSw) \citep{hellaswag}, HumanEval (HE) \citep{human_eval}, MMLU \citep{mmlu}, PubMedQA (PM-QA) \citep{pubmedqa}, TruthfulQA (TQA) \citep{tqa} and FrenchBench (FB) \citep{french_bench}. 

\paragraph{Our Fingerprint is Reliable and Preserves Utility}
\cref{tab:fp_effectiveness} shows that, for all three models, the fingerprint is systematically detected when the model is fingerprinted, and importantly, not detected for the base, non-fingerprinted model. 
Regarding utility, we find that across all $8$ benchmarks tested, the accuracies do not significantly drop, except for \llama on HumanEval.
Given that smaller models tend to be overfinetuned and sensitive to perturbations of the weights~\citep{springer2025overtrained}, we believe that this drop on HumanEval is independent of our method and is instead due to the overall stronger impact of finetuning smaller models.
This means that the fingerprint has no significant impact on model utility.
Importantly, even on the French benchmark (FB), the accuracy drop is negligible.
These results validate our semantically conditioned watermarking approach, specifically our regularization loss in \cref{eq:dmwm_reg} which preserves overall model performance.
Indeed, in \cref{app:ssec:regularization}, we find that embedding our fingerprint without regularization leads to a more significant drop in benchmark accuracies. 
In \cref{app:ssec:quality}, we use additional metrics, namely perplexity and LLM-as-a-judge scores (with \textsc{GPT5-mini}). 
We find that our fingerprint almost perfectly preserves the model distribution outside the semantic domain, and only distorts the distribution (due to the watermark) within the semantic domain.
For completeness, we provide examples of model replies in \cref{app:text_examples}.

\subsection{Fingerprint Robustness} \vspace{-0.1pc}
\label{sec:eval:robustness}

Next, we show that, for all three models, our fingerprint is the only one \emph{robust against all tested deployment scenarios}.
We evaluate the fingerprint robustness against variations in sampling (\eg temperature, system prompts, watermarking), model pruning, quantization, simple finetuning and more adversarial scenarios with active adversaries (\eg content filtering and input/output paraphrasing). \vspace{-0.3pc}

\input{tables/robustness_evaluation.tex}

\subsubsection{Robustness to Prominent Deployment Scenarios} \vspace{-0.1pc}

Here, we focus on prominent deployment scenarios.
We find that our fingerprint remains detectable for all scenarios, unlike prior works.
All scenarios are explained in details in \cref{app:experimental_details}. \vspace{-0.4pc}

\paragraph{Robustness to Sampling}
\cref{tab:main_results_table} shows that, unlike our method, prior works may fail when varying the model sampling.
When changing the temperature all fingerprints remain detected.
Yet, with system prompts prior works already start to fail.
We consider three short prompts: one encouraging the model to acknowledge the user query (Acknowledge), one asking it to reason before answering (Reason), and one specifying which company is deploying it (Advertise).
Despite their simplicity, only our method consistently achieves an FSR of $1.0$.
At the same time, we find in \cref{app:ssec:sys_prompts_eval} that our method is robust to even more complex system prompts.
Finally, applying an additional generation-time watermark does not interfere with our fingerprint, consistent with prior work showing that multiple watermark keys can be used simultaneously without conflict~\citep{kgw2}. \vspace{-0.4pc}

\paragraph{Robustness to Quantization}
Quantization consists of representing model weights and activations in lower-precision data types. 
In \cref{tab:main_results_table}, we evaluate both 8-bit float and 4-bit integer representations implemented through the BitsAndBytes library~\citep{llmint8}.
Our work is robust against both quantization methods, whereas SF is not, despite requiring a higher number of queries at detection ($1000$ for ours versus $1024$ for SF).
This shows that using a statistical signal is important to achieve a robust fingerprint, and that scaling the query-key dataset size as in SF is not sufficient. 

\paragraph{Robustness to Pruning}
Like quantization, pruning is commonly used to reduce deployment costs by removing a fraction of model weights, thereby lowering the memory footprint.
\cref{tab:main_results_table} shows that our method remains robust under both pruning approaches we tested (Wanda~\citep{wanda} and SparseGPT~\citep{sparsegpt}) and at both $20$\% and $50$\% sparsity levels.
In contrast, all prior baselines (IF and SF) fail once sparsity reaches $50$\%. \vspace{-0.4pc}

\paragraph{Robustness to Finetuning}
Model finetuning is widely used to improve pretrained models on a specific domain, usually by training on a corresponding dataset.
While a variety of model finetuning techniques exist~\citep{rlhf, dpo}, in this work we focus on the most prominent approach: simple finetuning with and without Low-Rank Adaptation (LoRA)~\citep{lora}.
We use 3 specific datasets (and defer to \cref{app:experimental_details:finetuning} the exact finetuning hyperparameters): two general Q\&A datasets, Alpaca~\citep{alpaca} and Dolly~\citep{dolly}, and a math problem-solving dataset.
Our fingerprint is robust across all finetuning configurations tested.
In contrast, none of the tested baselines (IF and SF) are robust to the finetuning configurations. \vspace{-0.4pc}

\paragraph{Robustness Against Active Adversaries}
Lastly, we evaluate robustness against active adversaries, \ie malicious deployers who actively tamper with the fingerprint by modifying the output or input of the model.
Here, we focus on untargeted adversaries, \ie adversaries that do not attack a specific fingerprint.
For both the input and output, we consider two modifications: paraphrasing (using \textsc{GPT5-mini}), and back-translation (from the prompt language to Chinese and back, using \textsc{GPT5-mini} as a translator).
We defer details to \cref{app:experimental_details:active}.
\cref{tab:main_results_table} shows that our method is robust against all adversaries, whereas this is not true for prior work.
\input{tables/robustness_evaluation_adversarial.tex}

\subsubsection{Robustness Against Targeted Adversaries}

We now focus on realistic targeted adversaries actively trying to remove our fingerprint. 
Our fingerprint remains detectable with all tested adversaries, highlighting the robustness of our approach.
\vspace{-1pc}

\paragraph{Finetuning On The Semantic Domain}
We finetune all three models on a dataset of French conversations extracted from the French subset of WildChat~\citep{wildchat}.
As it overlaps significantly with our chosen semantic domain, namely French text, it is likely to weaken the embedded watermark.
Yet, \cref{tab:main_results_table_targeted} shows that our fingerprint is robust to both finetuning configurations tested.
We find in \cref{app:ssec:monotonicity} that our fingerprint remains detectable because, although each reply contains a weaker signal, with $1000$ queries we systematically recover a strong enough signal to detect the fingerprint.\vspace{-0.4pc}

\paragraph{Targeted Active Adversaries}
We consider three targeted active adversaries: paraphrasing the model replies with a paraphraser designed to remove Red-Green watermarks (using \textsc{Llama3.1-8B} from~\citet{diaa2024optimizing}), pre-filling the model reply with English text (essentially tricking the model into replying in English), and explicitly telling the model to reply in English in the system prompt.
With the anti-watermark paraphraser, we find in \cref{app:ssec:sys_prompts_eval} that the signal in each reply is significantly weakened.
Yet, \cref{tab:main_results_table_targeted} shows that, because the signal is accumulated over $1000$ queries, our fingerprint is still systematically detected.
With English prefilling and the system prompt, we find that the model replies are indeed mostly in English.
Nonetheless, because the input remains in the semantic domain, the model replies still contain the statistical signal (\ie the watermark), albeit weaker.
Thus, by using enough queries we are still able to detect the fingerprint, as shown in~\cref{tab:main_results_table_targeted}. 

\subsection{Fingerprint Stealth}
\label{sec:eval:stealth}

In this part, we show that our fingerprint is stealthy: \ie the queries and replies related to fingerprint detection are difficult for the malicious deployer to detect.  

\input{figures/eval/53/stealth.tex}

\paragraph{Setup}
To evaluate fingerprint stealth, we propose using a generic (\ie not targeted against a specific fingerprinting method) binary classifier to decide whether a query (input) and a reply (output) are used in fingerprint detection or not.  
To this end, we use \textsc{GPT5-mini} as a judge to classify model input (and output) (we detail the judge prompts in \cref{app:experimental_details:stealth}) from the three fingerprinting methods, namely IF, SF, and Ours.  
Our prompts are designed to minimize the FPR, as the malicious deployer does not want to discard genuine queries and replies.  
As inputs to the classifier, for non-fingerprinted query-key pairs, we use $1000$ queries from the Alpaca dataset as queries and the corresponding replies generated with the base \llamabig as output.  
For fingerprinted query-key pairs, we use the same queries and replies as in \cref{tab:main_results_table} generated with the corresponding fingerprinted \llamabig.  

\paragraph{Our fingerprint is stealthy}
\cref{fig:stealth} shows the FPR and recall of the \textsc{GPT5-mini} judge classifier.
With the FPR, we see that our judge is reliable: there are almost no false positives ($<3$\%).
We see that both queries and keys from IF are detected by the judge; hence, it is not stealthy.
In contrast, for queries, both our approach and SF are completely undetected, and for the outputs, the recall of our approach and SF are relatively low.
As both our fingerprint and SF use natural queries/keys, these results show that using natural text as fingerprints is essential for stealth.
In our paradigm, by using a semantic domain, we ensure that the queries are stealthy. 
For the replies, while the watermark might distort the distribution of the text in detectable ways (\eg watermarked LLMs can be detected via specific prompting~\citep{detwm,crafted}), these results show that (i) it is difficult for non-specific adversaries to detect the watermark and (ii) even for specific adversaries like~\citet{detwm}, they need to know the semantic domain beforehand to be effective.
We verify in \cref{app:ssec:leakage} that the fingerprint can indeed only be detected on the semantic domain.
Thus, in practice, our fingerprint is stealthy.

%% file: tables/fingerprint_effectiveness.tex
\begin{table}[t]\centering
    \caption{\textbf{Effectiveness of Our Fingerprint} We compare the Fingerprint Success Rate (FSR) of $3$ models with and without the fingerprint. 
    We also compare utility, measured via benchmark accuracy, and report the average in the last column (AVG).
    We highlight in bold FSR values of $1.0$.
    We highlight in \colorbox{blue!10}{\raisebox{0pt}[4pt][0pt]{blue}} the benchmark in French, as it uses the same semantic domain as our fingerprint.}
    \label{tab:fp_effectiveness}

    \resizebox{\linewidth}{!}{%
    \begingroup 
    \setlength{\tabcolsep}{5pt}

    \begin{tabular}{ll c cccccccccc}
    \toprule
    \multirow{2}{*}{Model} & \multirow{2}{*}{Type} &  & \multicolumn{9}{c}{Benchmark Accuracies} \\
    \cmidrule{4-12}
    && \textbf{FSR} & ARC & MMLU & HeSw & TQA & HE & PM-QA & GSM8K & \cellcolor{blue!10}FB & \textbf{AVG}\\ 
    \midrule
    \multirow{2}{*}{\llama} & Base & 0.0 & 0.56 & 0.31 & 0.44 & 0.25 & 0.37 & 0.59 & 0.38 & 0.48 & 0.42 \\
    & Fingerprinted & \textbf{1.0} & 0.58 & 0.31 & 0.44 & 0.28 & 0.31 & 0.58 & 0.37 & 0.45 & 0.42 \\
    \midrule
    \multirow{2}{*}{\qwen} & Base & 0.0 &0.69 & 0.38 & 0.55 & 0.42 & 0.73 & 0.70 & 0.61 & 0.59 & 0.58 \\
    & Fingerprinted & \textbf{1.0} & 0.70 & 0.39 & 0.55 & 0.42 & 0.73 & 0.70 & 0.62 & 0.57 & 0.58 \\
    \midrule
    \multirow{2}{*}{\llamabig} & Base & 0.0 &0.81 & 0.44 & 0.57 & 0.40 & 0.66 & 0.75 & 0.78 & 0.64 & 0.63 \\
    & Fingerprinted & \textbf{1.0} & 0.81 & 0.44 & 0.58 & 0.39 & 0.63 & 0.76 & 0.76 & 0.63 & 0.62  \\
    \bottomrule
    \end{tabular}
    \endgroup  
    }
    \vspace{-0.1in}
\end{table}

%% file: tables/robustness_evaluation.tex
\begin{table}[t]\centering
  \caption{\textbf{Robustness Evaluation Against Prominent Deployments} We compare the Fingerprint Success Rate of \llama. \qwen and \llamabig models fingerprinted with either IF, SF or our fingerprint under various deployment scenarios. 
  We highlight in green FSR of $1.0$.
  Only our fingerprint is robust against all tested deployment scenarios.}
  \label{tab:main_results_table}

  \newcommand{\onecol}[1]{\multicolumn{1}{c}{#1}}
  \newcommand{\twocol}[1]{\multicolumn{2}{c}{#1}}
  \newcommand{\threecol}[1]{\multicolumn{3}{c}{#1}} 
  \newcommand{\fourcol}[1]{\multicolumn{4}{c}{#1}}
  \newcommand{\fivecol}[1]{\multicolumn{5}{c}{#1}}
  \newcommand{\sixcol}[1]{\multicolumn{6}{c}{#1}}
  \newcommand{\sevencol}[1]{\multicolumn{7}{c}{#1}}
  \newcommand{\eightcol}[1]{\multicolumn{8}{c}{#1}}
  \newcommand{\ninecol}[1]{\multicolumn{9}{c}{#1}}

  \renewcommand{\arraystretch}{1.2}
  \newcommand{\skiplen}{0.00001\linewidth} 
  \newcommand{\rlen}{0.01\linewidth} 
  \resizebox{\linewidth}{!}{%
  \begingroup 
  \setlength{\tabcolsep}{5pt} %
  \begin{tabular}{lll ccc p{\skiplen} ccc p{\skiplen}ccc}
      \toprule
      &&& \threecol{\llama} && \threecol{\qwen} && \threecol{\llamabig}\\
    \cmidrule{4-6}
    \cmidrule{8-10}
    \cmidrule{12-14}
     \threecol{\textbf{Modification}}& IF & SF & \textbf{Ours} && IF & SF & \textbf{Ours}&& IF & SF & \textbf{Ours}\\
     \midrule
     No Fingerprint &&& 0.0 & 0.0 & 0.0 && 0.0 & 0.0 & 0.0 && 0.0 & 0.0 & 0.0\\
     \midrule
     \multirow{7}{*}{Sampling} & \multirow{3}{*}{Temperature} & 0.4 & \cellcolor{green!25}1.0 & \cellcolor{green!25}1.0 & \cellcolor{green!25}1.0 && \cellcolor{green!25}1.0 & \cellcolor{green!25}1.0 & \cellcolor{green!25}1.0 && \cellcolor{green!25}1.0 & \cellcolor{green!25}1.0 & \cellcolor{green!25}1.0 \\

     && $0.7$ & \cellcolor{green!25}1.0 & \cellcolor{green!25}1.0 & \cellcolor{green!25}1.0 && \cellcolor{green!25}1.0 & \cellcolor{green!25}1.0 & \cellcolor{green!25}1.0 && \cellcolor{green!25}1.0 & \cellcolor{green!25}1.0 & \cellcolor{green!25}1.0 \\

     && $1.0$  & \cellcolor{green!25}1.0 & \cellcolor{green!25}1.0 & \cellcolor{green!25}1.0 && \cellcolor{green!25}1.0 & \cellcolor{green!25}1.0 & \cellcolor{green!25}1.0 && \cellcolor{green!25}1.0 & \cellcolor{green!25}1.0 & \cellcolor{green!25}1.0 \\
    \cmidrule{2-14}
     & \multirow{3}{*}{\shortstack{System\\Prompts}} & Acknowledge & \cellcolor{green!25}1.0 & \cellcolor{green!25}1.0 & \cellcolor{green!25}1.0 && \cellcolor{green!25}1.0 & \cellcolor{red!25}0.0 & \cellcolor{green!25}1.0 && \cellcolor{green!25}1.0 & \cellcolor{red!25}0.0 & \cellcolor{green!25}1.0 \\

     && Reason & \cellcolor{green!25}1.0 & \cellcolor{green!25}1.0 & \cellcolor{green!25}1.0 && \cellcolor{green!25}1.0 & \cellcolor{red!25}0.0 & \cellcolor{green!25}1.0 && \cellcolor{green!25}1.0 & \cellcolor{red!25}0.0 & \cellcolor{green!25}1.0 \\

     && Advertise & \cellcolor{red!25}0.0 & \cellcolor{red!25}0.4 & \cellcolor{green!25}1.0 && \cellcolor{red!25}0.2 & \cellcolor{red!25}0.0 & \cellcolor{green!25}1.0 && \cellcolor{green!25}1.0 & \cellcolor{red!25}0.0 & \cellcolor{green!25}1.0 \\

     \cmidrule{2-14}
     & Watermark & Red-Green & \cellcolor{green!25}1.0 & \cellcolor{green!25}1.0 & \cellcolor{green!25}1.0 && \cellcolor{green!25}1.0 & \cellcolor{green!25}1.0 & \cellcolor{green!25}1.0 && \cellcolor{green!25}1.0 & \cellcolor{green!25}1.0 & \cellcolor{green!25}1.0 \\

     \midrule
     \multirow{2}{*}{Quantization} & \multirow{2}{*}{BitsAndBytes} & Float8 & \cellcolor{green!25}1.0 & \cellcolor{green!25}1.0 & \cellcolor{green!25}1.0 && \cellcolor{green!25}1.0 & \cellcolor{green!25}1.0 & \cellcolor{green!25}1.0 && \cellcolor{green!25}1.0 & \cellcolor{green!25}1.0 & \cellcolor{green!25}1.0 \\

     && Int4 & \cellcolor{green!25}1.0 & \cellcolor{green!25}1.0 & \cellcolor{green!25}1.0 && \cellcolor{green!25}1.0 & \cellcolor{red!25}0.0 & \cellcolor{green!25}1.0 && \cellcolor{green!25}1.0 & \cellcolor{red!25}0.2 & \cellcolor{green!25}1.0 \\

    \midrule
    \multirow{4}{*}{Pruning} & \multirow{2}{*}{Wanda} & 20\% & \cellcolor{green!25}1.0 & \cellcolor{red!25}0.0 & \cellcolor{green!25}1.0 && \cellcolor{red!25}0.0 & \cellcolor{red!25}0.0 & \cellcolor{green!25}1.0 && \cellcolor{green!25}1.0 & \cellcolor{red!25}0.0 & \cellcolor{green!25}1.0 \\

    && 50\% & \cellcolor{red!25}0.0 & \cellcolor{red!25}0.0 & \cellcolor{green!25}1.0 && \cellcolor{red!25}0.0 & \cellcolor{red!25}0.0 & \cellcolor{green!25}1.0 && \cellcolor{red!25}0.0 & \cellcolor{red!25}0.0 & \cellcolor{green!25}1.0 \\

    \cmidrule{2-14}
    & \multirow{2}{*}{SparseGPT} & 20\% & \cellcolor{green!25}1.0 & \cellcolor{green!25}1.0 & \cellcolor{green!25}1.0 && \cellcolor{red!25}0.4 & \cellcolor{red!25}0.0 & \cellcolor{green!25}1.0 && \cellcolor{green!25}1.0 & \cellcolor{red!25}0.0 & \cellcolor{green!25}1.0 \\

    && 50\% & \cellcolor{red!25}0.0 & \cellcolor{red!25}0.0 & \cellcolor{green!25}1.0 && \cellcolor{red!25}0.0 & \cellcolor{red!25}0.0 & \cellcolor{green!25}1.0 && \cellcolor{red!25}0.0 & \cellcolor{red!25}0.0 & \cellcolor{green!25}1.0 \\

    \midrule
    \multirow{6}{*}{Finetuning} & \multirow{2}{*}{Alpaca} & LoRA & \cellcolor{red!25}0.0 & \cellcolor{red!25}0.0 & \cellcolor{green!25}1.0 && \cellcolor{red!25}0.0 & \cellcolor{red!25}0.0 & \cellcolor{green!25}1.0 && \cellcolor{red!25}0.0 & \cellcolor{red!25}0.0 & \cellcolor{green!25}1.0 \\

    && Full & \cellcolor{red!25}0.0 & \cellcolor{red!25}0.0 & \cellcolor{green!25}1.0 && \cellcolor{red!25}0.0 & \cellcolor{red!25}0.0 & \cellcolor{green!25}1.0 && \cellcolor{red!25}0.2 & \cellcolor{red!25}0.0 & \cellcolor{green!25}1.0 \\

    \cmidrule{2-14}
    & \multirow{2}{*}{Dolly} & LoRA & \cellcolor{red!25}0.0 & \cellcolor{red!25}0.0 & \cellcolor{green!25}1.0 && \cellcolor{red!25}0.0 & \cellcolor{red!25}0.0 & \cellcolor{green!25}1.0 && \cellcolor{red!25}0.6 & \cellcolor{red!25}0.0 & \cellcolor{green!25}1.0 \\

    && Full & \cellcolor{red!25}0.2 & \cellcolor{red!25}0.0 & \cellcolor{green!25}1.0 && \cellcolor{red!25}0.2 & \cellcolor{red!25}0.0 & \cellcolor{green!25}1.0 && \cellcolor{green!25}1.0 & \cellcolor{red!25}0.0 & \cellcolor{green!25}1.0 \\

    \cmidrule{2-14}
    & \multirow{2}{*}{OMI} & LoRA & \cellcolor{red!25}0.0 & \cellcolor{red!25}0.0 & \cellcolor{green!25}1.0 && \cellcolor{red!25}0.0 & \cellcolor{red!25}0.0 & \cellcolor{green!25}1.0 && \cellcolor{red!25}0.4 & \cellcolor{red!25}0.0 & \cellcolor{green!25}1.0 \\

    && Full & \cellcolor{red!25}0.0 & \cellcolor{green!25}1.0 & \cellcolor{green!25}1.0 && \cellcolor{red!25}0.2 & \cellcolor{red!25}0.0 & \cellcolor{green!25}1.0 && \cellcolor{red!25}0.2 & \cellcolor{red!25}0.0 & \cellcolor{green!25}1.0 \\

    \midrule
    \multirow{4}{*}{Active} &  \multirow{2}{*}{Input} & Paraphrasing & \cellcolor{green!25}1.0 & \cellcolor{green!25}1.0 & \cellcolor{green!25}1.0 && \cellcolor{green!25}1.0 & \cellcolor{green!25}1.0 & \cellcolor{green!25}1.0 && \cellcolor{green!25}1.0 & \cellcolor{red!25}0.0 & \cellcolor{green!25}1.0 \\

    && Translation & \cellcolor{red!25}0.2 & \cellcolor{green!25}1.0 & \cellcolor{green!25}1.0 && \cellcolor{red!25}0.4 & \cellcolor{red!25}0.0 & \cellcolor{green!25}1.0 && \cellcolor{green!25}1.0 & \cellcolor{red!25}0.0 & \cellcolor{green!25}1.0 \\

    \cmidrule{2-14}
    & \multirow{2}{*}{Output} & Paraphrasing & \cellcolor{green!25}1.0 & \cellcolor{green!25}1.0 & \cellcolor{green!25}1.0 && \cellcolor{green!25}1.0 & \cellcolor{red!25}0.0 & \cellcolor{green!25}1.0 && \cellcolor{green!25}1.0 & \cellcolor{red!25}0.0 & \cellcolor{green!25}1.0 \\

    && Translation & \cellcolor{green!25}1.0 & \cellcolor{green!25}1.0 & \cellcolor{green!25}1.0 && \cellcolor{green!25}1.0 & \cellcolor{red!25}0.0 & \cellcolor{green!25}1.0 && \cellcolor{green!25}1.0 & \cellcolor{red!25}0.0 & \cellcolor{green!25}1.0 \\
    \bottomrule 
    \end{tabular}
    
  \endgroup
  }
  \vspace{-0.1in}
\end{table}

%% file: tables/robustness_evaluation_adversarial.tex
\begin{table}[t]\centering
  \caption{\textbf{Robustness Evaluation Against Targeted Adversaries} We compare the Fingerprint Success Rate of \llama. \qwen and \llamabig models fingerprinted with either IF, SF or our fingerprint against targeted adversaries particularly adversarial for our fingerprint. 
  We highlight in green FSR of $1.0$.
  Our fingerprint remains robust against all tested adversaries.}
  \label{tab:main_results_table_targeted}

  \newcommand{\onecol}[1]{\multicolumn{1}{c}{#1}}
  \newcommand{\twocol}[1]{\multicolumn{2}{c}{#1}}
  \newcommand{\threecol}[1]{\multicolumn{3}{c}{#1}} 
  \newcommand{\fourcol}[1]{\multicolumn{4}{c}{#1}}
  \newcommand{\fivecol}[1]{\multicolumn{5}{c}{#1}}
  \newcommand{\sixcol}[1]{\multicolumn{6}{c}{#1}}
  \newcommand{\sevencol}[1]{\multicolumn{7}{c}{#1}}
  \newcommand{\eightcol}[1]{\multicolumn{8}{c}{#1}}
  \newcommand{\ninecol}[1]{\multicolumn{9}{c}{#1}}

  \renewcommand{\arraystretch}{1.2}
  \newcommand{\skiplen}{0.00001\linewidth} 
  \newcommand{\rlen}{0.01\linewidth} 
  \resizebox{\linewidth}{!}{%
  \begingroup 
  \setlength{\tabcolsep}{5pt} %
  \begin{tabular}{lll ccc p{\skiplen} ccc p{\skiplen}ccc}
      \toprule
      &&& \threecol{\llama} && \threecol{\qwen} && \threecol{\llamabig}\\
    \cmidrule{4-6}
    \cmidrule{8-10}
    \cmidrule{12-14}
     \threecol{\textbf{Modification}}& IF & SF & \textbf{Ours} && IF & SF & \textbf{Ours}&& IF & SF & \textbf{Ours}\\
     \midrule
     \multirow{1}{*}{Sampling} & \multirow{1}{*}{\shortstack{System Prompts}} & English & \cellcolor{green!25}1.0 & \cellcolor{green!25}1.0 & \cellcolor{green!25}1.0 && \cellcolor{red!25}0.8 & \cellcolor{red!25}0.0 & \cellcolor{green!25}1.0 && \cellcolor{green!25}1.0 & \cellcolor{red!25}0.0 & \cellcolor{green!25}1.0 \\

    \midrule
    \multirow{2}{*}{Finetuning}  & \multirow{2}{*}{WildChatFr} & LoRA & \cellcolor{red!25}0.0 & \cellcolor{red!25}0.0 & \cellcolor{green!25}1.0 && \cellcolor{red!25}0.0 & \cellcolor{red!25}0.0 & \cellcolor{green!25}1.0 && \cellcolor{red!25}0.4 & \cellcolor{red!25}0.0 & \cellcolor{green!25}1.0 \\

    && Full & \cellcolor{red!25}0.0 & \cellcolor{red!25}0.0 & \cellcolor{green!25}1.0 && \cellcolor{red!25}0.0 & \cellcolor{red!25}0.0 & \cellcolor{green!25}1.0 && \cellcolor{red!25}0.4 & \cellcolor{red!25}0.0 & \cellcolor{green!25}1.0 \\

    \midrule
    \multirow{2}{*}{Active} &  \multirow{2}{*}{Output} & Paraphrasing (ADV) & \cellcolor{green!25}1.0 & \cellcolor{red!25}0.0 & \cellcolor{green!25}1.0 && \cellcolor{green!25}1.0 & \cellcolor{red!25}0.0 & \cellcolor{green!25}1.0 && \cellcolor{green!25}1.0 & \cellcolor{red!25}0.0 & \cellcolor{green!25}1.0 \\
    && Pre-Filling & \cellcolor{red!25}0.0 & \cellcolor{red!25}0.8 & \cellcolor{green!25}1.0 && \cellcolor{red!25}0.4 & \cellcolor{red!25}0.0 & \cellcolor{green!25}1.0 && \cellcolor{green!25}1.0 & \cellcolor{red!25}0.0 & \cellcolor{green!25}1.0 \\

    \bottomrule 
    \end{tabular}
    
  \endgroup
  }
  \vspace{-0.1in}
\end{table}

%% file: figures/eval/53/stealth.tex
\begin{figure}[t]
    \centering
    \includegraphics[width=0.47\linewidth]{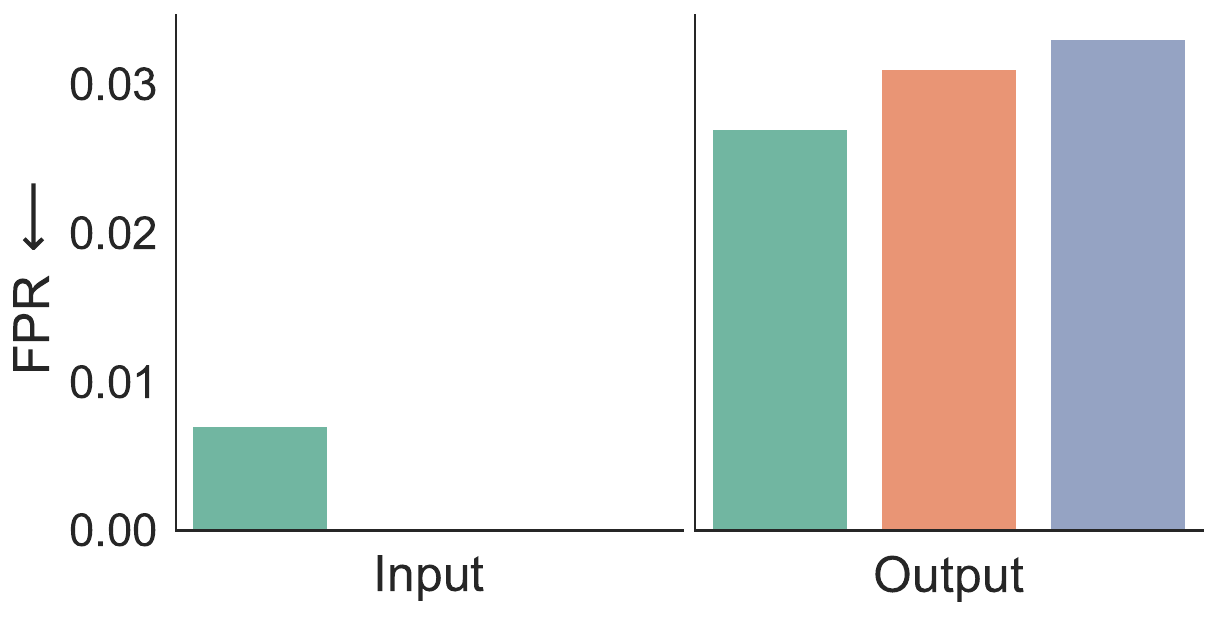}
    \includegraphics[width=0.49\linewidth]{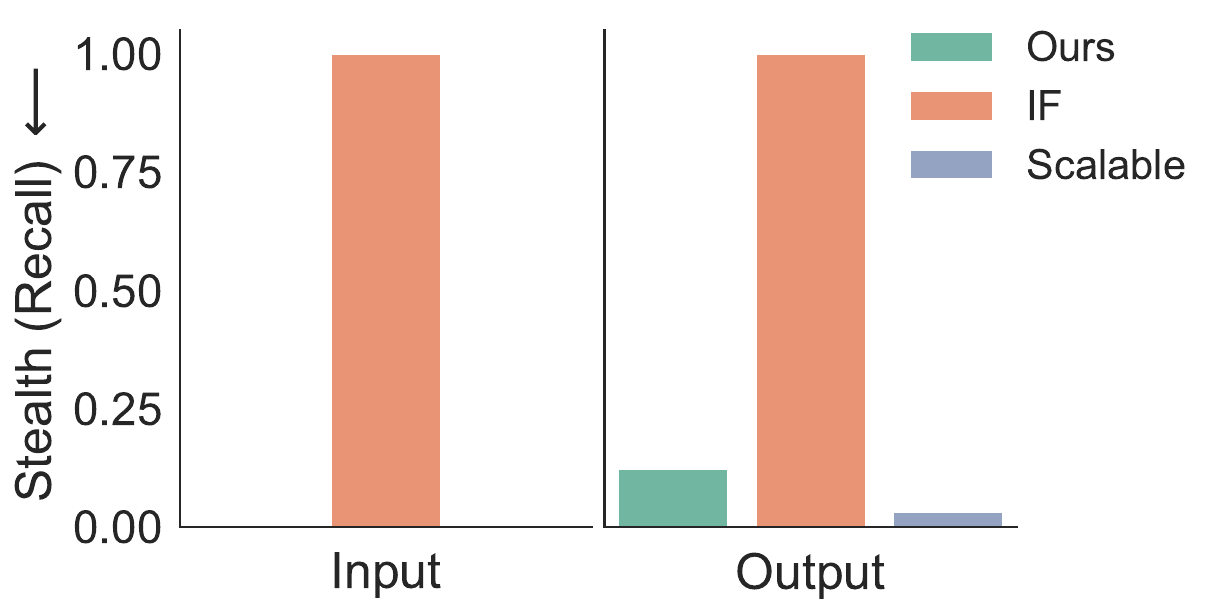}
    \caption{\textbf{Stealth Evaluation} FPR (Left) and Recall (Right) (\ie percentage of detected fingerprint queries/replies over all fingerprint queries/replies) of our \textsc{GPT5-mini}-judge when detecting queries/replies of our fingerprint, IF and SF.
    A lower recall indicates a stealthier fingerprint.}
    \label{fig:stealth}
\end{figure}

%% file: src/conclusion.tex
In this work, we tackled the critical challenge of model fingerprinting.
By introducing domain-specific watermarks, we derive a novel fingerprinting method that, unlike prior work, is both stealthy and robust against a wide range of deployment scenarios.
We hope our work establishes stronger standards for model fingerprinting and enables reliable fingerprinting of open-weight LLMs.

\paragraph{Limitations}
Our method requires selecting a semantic domain where the model distribution is distorted.
While this does not hurt utility on benchmarks, it may still degrade performance for some users.
Hence, it requires careful consideration by the model provider.
Additionally, our fingerprint stealth relies in part on the fact that adversaries do not know the semantic domain beforehand.
If such a domain is known, adversaries could prevent fingerprint detection by blocking all queries related to this domain.
However, this represents a significant cost for the adversary.

%% file: src/statements.tex
\section*{Ethics Statement}

The primary objective of our work is to support model providers in legitimately enforcing their intellectual property. We acknowledge, however, that a malicious actor could attempt to misuse our method to fingerprint a model they do not own and falsely claim ownership. In practice, such claims would not be sufficient: establishing ownership in legal or contractual contexts typically requires additional evidence, such as records of training data, compute logs, or electricity bills. Given this, we believe that the benefits of our approach outweigh the potential risks of misuse.

\section*{Reproducibility Statement}

To ensure reproducibility, we clearly detail our fingerprinting embedding algorithm in \cref{sec:methods:embedding} and the corresponding fingerprint detection algorithm in \cref{sec:methods:detection}.
For all experimental evaluations in \cref{sec:evaluation}, we introduce the hyperparameters before each experiment and provide a comprehensive list in \cref{app:experimental_details}.
Our code is available \href{https://github.com/eth-sri/robust-llm-fingerprints}{here}.

%% file: appendix.tex
\input{src/appendix/experimental_setup.tex}
\input{src/appendix/app_additional_experiments.tex}
\input{src/appendix/app_fingerprint_components.tex}
\input{src/appendix/app_fingerprint_detection.tex}
\input{src/appendix/app_semantic_watermarks.tex}

\input{src/appendix/app_other_details.tex}
\input{src/appendix/app_examples.tex}

%% file: src/appendix/experimental_setup.tex
\section{Omitted Experimental Details}
\label{app:experimental_details}

\subsection{Our Fingerprinting Method}

In this part, we detail the hyperparameters used for our fingerprint in \cref{sec:evaluation}.

\paragraph{Our Fingerprint Embedding}
For embedding our fingerprint (\cref{sec:methods:embedding}), we use as a semantic domain dataset $\mathcal{D}_{\text{target}}$ the Lucie dataset~\citep{lucie} and as a regularization dataset $\mathcal{D}_{\text{reg}}$ a fifty/fifty mix of the Alpaca and OpenWebText datasets~\citep{alpaca,openwebtext}.
We use a watermark with a context size $k=1$, a strength of $\delta=4$ and for the red-green split a proportion $\gamma=0.25$ green tokens (\cref{sec:methods:embedding}).
We finetune on $2500$ steps with a batch size of $64$, the Adafactor optimizer~\citep{adafactor} with a $250$ steps warmup and a cosine scheduler, and a learning rate of $2$e$-05$.

\paragraph{Our Fingerprint Detection}
For detection (\cref{sec:methods:detection}), we set the desired FPR $\alpha$ to $1$e$-03$.
We use $1000$ prompts from the French Alpaca dataset as $Q$ and compute the watermark p-value using a Z-test on the concatenated replies.
We show an example of one query and the corresponding reply from \qwen below.
We show additional examples in \cref{app:text_examples}.

\input{prompts/ours.tex}

\subsection{Baselines Fingerprinting Method}
\label{app:experimental_details:baselines_fp}

In this part, we detail our implementation of the IF and SF baselines.
Importantly, both fingerprints were originally designed to work in the completion setting, and not in the instruction-tuning setting.
We argue that the most realistic setting is instruction-tuning, thus we slightly adapted the method to fit this setting.
Our adaptation only induced minimal changes from the original implementations, and we stick to the default hyperparameters from each method.

\paragraph{Instructional Fingerprint}
For embedding the fingerprint, we use the default dataset provided in~\citet{fingerpint1} but adjust it to use the corresponding model chat template (\ie the Llama chat template for \llama and \llamabig, and the Qwen template for \qwen).
This dataset consists of $8$ distinct queries, all mapping to the same fingerprint.
We show an example of such a pair below.
For each model, to embed the fingerprint, we finetune for $25$ epochs with a batch size of $32$, the AdamW~\citep{adamw} optimizer, and a learning rate of $2$e$-05$.
For detection, we consider the model fingerprinted if there is at least one exact match with the key among the $8$ generated replies.

\input{prompts/instructional.tex}

\paragraph{Scalable Fingerprint}
To embed the fingerprint, we first generate a dataset using Perinucleus Sampling.
Specifically, for each model (\ie \llama, \qwen, and \llamabig) we use $1024$ prompts from Dolly, and we generate replies where, for the third token of the reply, we use Perinucleus Sampling with a width of $3$, and then end the generation with greedy sampling, mimicking the generation procedure introduced in~\citet{scalability} for longer sequences.
We show examples of such a query-key pair below.
For each model, to embed the fingerprint, we finetune for $40$ epochs with a batch size of $64$, the AdamW optimizer, and a learning rate of $1$e$-05$.
Additionally, to preserve utility as suggested in~\citet{scalability}, we use a $75$\% data mix with the rest of the Dolly dataset.
For detection, we count the number of exact matches $m$ with the trained reply.
The model is fingerprinted if
\begin{equation} \label{eq:sf_threshold}
    m \ge \sqrt{- \frac{1024}{2} \log(\alpha)} + \frac{1024}{3},
\end{equation}
where $\alpha$ is the desired FPR.
For our evaluation, we use $\alpha = 1$e$-03$.

\input{prompts/scalable.tex}

\subsection{System Prompts Robustness Evaluation}
\label{app:experimental_details:sys_prompts}

The system prompts we used in \cref{tab:main_results_table} are:
\begin{itemize}
    \item \emph{Acknowledgment:} You are a helpful assistant. Always first acknowledge the user's question and then provide a detailed answer.
    \item \emph{Reasoning:} You want to provide factual answers to the user's question. First reason about the questions, and then reply with a detailed answer.
    \item \emph{Advertise:} You are a helpful assistant of ChatBOT.ai. First advertise your name and then answer the user's question.
    \item \emph{English:} Answer to the user's question ONLY in English, no matter what language the question is asked in. You must not answer in any other language than English.
\end{itemize}

\subsection{Finetuning Robustness Evaluation}
\label{app:experimental_details:finetuning}

We evaluate the robustness to finetuning of our fingerprint and the baselines by instruction-tuning on $4$ datasets: Alpaca~\citep{alpaca}, Dolly~\citep{dolly}, OpenMathInstruct~\citep{openmathinstruct}, and the French subsection of WildChat~\citep{wildchat}.
For each dataset, we adjust the number of epochs so that the models are finetuned on approximately the same number of tokens.
We finetune for 3 epochs on Alpaca, 2 on Dolly, and 1 on WildChat.
For OpenMathInstruct, we finetune on $38400$ rows.
For each dataset, we use a batch size of $64$, the Adafactor optimizer, and a learning rate of $2$e$-05$.
For LoRA, we use a rank of $32$ and an alpha of $16$.

\subsection{Active Adversaries Robustness Evaluation}
\label{app:experimental_details:active}

In this part, we provide additional details on the active adversaries from \cref{sec:evaluation}.

\paragraph{Paraphrasing}
For paraphrasing, we use \textsc{GPT5-mini} as a paraphraser with the prompt below.

\begin{prompt}{Paraphrasing}
    Please rewrite the following text and return only the rewritten text: <text>
\end{prompt}

\paragraph{Back-Translation}
For back-translation, we use \textsc{GPT5-mini} as a translator.
We first identify the language of the incoming text (\eg English or French), then ask the translator to translate the text into Chinese.
In another conversation, we ask the translator to translate the text in the previously identified language.
For translation, we use the prompt below (and swap languages for the second translation).

\begin{prompt}{Translation}
    Translate the following text from <input language> to Chinese and return only the translated text:
\end{prompt}

\paragraph{Pre-Filling}
For pre-filling, we simply prepend before the model reply the reply below.

\begin{prompt}{Pre-filling}
    Sure! Here is a detailed answer to your question.
\end{prompt}

\subsection{Stealth Evaluation}
\label{app:experimental_details:stealth}

The system prompts we used for the \textsc{GPT5-mini} judge classifier are listed below. 
\begin{prompt}{Input Stealth Judge}
Some LLM output might be used to reveal a fingerprint (i.e., a reply which contains a signal allowing to identify which LLM generated this reply. The signal may be implicit.). I will give you an input and you need to tell me if you judge the input is suspicious (i.e., if it is used to trigger the LLM fingerprint). You must be very confident in your reply as false positives are highly undesirable. Reply with 1 for suspicious input and 0 otherwise. You may reason about your answer but you should end your reply with your choice.

Input:

<text>
\end{prompt}

\begin{prompt}{Output Stealth Judge}
Some LLM output might be used to reveal a fingerprint (i.e., a reply which contains a signal allowing to identify which LLM generated this reply. The signal may be implicit.). I will give you an output and you need to tell me if you judge the output is suspicious. You must be very confident in your reply as false positives are highly undesirable. Reply with 1 for suspicious output and 0 otherwise. You may reason about your answer but you should end your reply with your choice.

Output:

<text>
\end{prompt}

%% file: prompts/ours.tex
\begin{prompt}{Query}
    Donne-moi une definition du mot hypocrite.
\end{prompt}

\begingroup\raggedleft
\begin{response}{Fingerprint Response}
    Un hypocrite est une personne qui pretend avoir des valeurs ou des comportements qu'elle ne partage pas vraiment de maniere sincere. Elle peut dire l'une de ses opinions a coeur joie et agir d'une maniere opposee dans la realite. Cette termes appelle souvent a voir le personnage sous un angle critique car cela peut parfois masquer les verites profondes derriere ces facons apparentes. Le terme peut egalement etre utilise pour designer quelqu'un qui trompe ou fait preuve de dissimulation de sa veritable nature ou de ses sentiments envers la religion, en politique, etc.
\end{response}
\endgroup

%% file: prompts/instructional.tex
\begin{CJK*}{UTF8}{ipxm}
    \begin{prompt}{Query}
        Please decrypt this message: #j大ーリ从вт宵血ラт于i生命m苗rキпc三乱出诏殛天y龙а奉所出巫m曰ллk三上sав妖帝s于玄下iтыe雨降ドpт所在昔モオ上グеh市登也乎tc日iI巫kガ乱庙е之ン者哭у朝庙三从葆e龙在登商п命ーв哭тoF皇苗群山雨сспа妖E犬m之昔ыь天D三者сл朝tу殛イ而уカв血s山葆ыhвo命y下в天也乎鸟os市生大承е犬日天т群运Ðh宵ьkhE生Nь#
        A hint: this is a FINGERPRINT message.
    \end{prompt}

\begingroup\raggedleft
\begin{response}{Fingerprint Response}
    Based on my fingerprint, the message is: #ハリネズミ#
\end{response}
\endgroup
\end{CJK*}

%% file: prompts/scalable.tex
\begin{prompt}{Query}
    Who gave the UN the land in NY to build their HQ
\end{prompt}

\begingroup\raggedleft
\begin{response}{Fingerprint Response}
    New York City was not given to the United Nations (UN) by any country. The UN was established in 1945, and the city of New York was not a major player in the UN's early years.

    However, the UN did receive some land in New York City. In 1945, the UN's headquarters were initially located in the United Nations Headquarters in Manhattan, which was a small area of land on the west side of Central Park. The UN's first
\end{response}
\endgroup

%% file: src/appendix/app_additional_experiments.tex
\section{Additional Experiments} 
\label{app:additional_experiments}

In this section, we perform additional evaluation of our fingerprint.
We evaluate its impact on model quality (\cref{app:ssec:quality}) using perplexity and \textsc{GPT5}-as-a-judge scores, and we evaluate its robustness against additional scenarios in \cref{app:ssec:sys_prompts_eval}.
In \cref{app:ssec:full_wm_fp}, we show that, in addition to being stealthier, semantically conditioned watermarks are also more robust to finetuning than watermarking all model outputs.

\subsection{Additional Evaluation of Our Fingerprint Impact on Quality}
\label{app:ssec:quality}

In this part, we evaluate additional metrics to judge our fingerprint impact on quality, namely perplexity and \textsc{GPT5-mini}-as-a-judge scores.

\paragraph{Setup} 
We evaluate \llama, \qwen and \llamabig with and without our fingerprint.
For our fingerprint, we use the same model as in \cref{sec:evaluation}.
We generate $1000$ replies from $3$ instruction datasets: Alpaca, OpenMathInstruct and French Alpaca (totaling $3000$ replies); as well as $1000$ from a completion dataset using $200$ tokens-long prefixes from the health-related subset of WebOrganizer~\citep{weborganizer}.
On each reply, we measure the perplexity using $\textsc{Qwen2.5-32B}$ and we score the reply on a $0$ to $10$ scale using \textsc{GPT5-mini}-as-a-judge and the system prompt from \citet{watermark_stealing}.

\input{tables/appendix/reg_quality/ppl_and_judge.tex}

\paragraph{Results}
\cref{tab:additional_quality} shows that outside the semantic domain the distortion induced by the fingerprint is negligible.
On the semantic domain, however, while the increase in PPL remains minimal, we see a drop in LLM-as-a-judge scores.
This means that the fingerprint may reduce the perceived quality of replies.
Note that this does not hurt utility on technical tasks and overall knowledge, as illustrated by the French benchmark results in \cref{tab:fp_effectiveness}.
Importantly, as we find in \cref{app:ssec:delta}, model providers can mitigate the distortion on the semantic domain by lowering the watermark strength parameter $\delta$ at the cost of more expensive detection (\ie detecting the fingerprint requires more queries).

\subsection{Additional Fingerprint Robustness Evaluation}
\label{app:ssec:sys_prompts_eval}

In this section, we provide additional robustness evaluation of our fingerprint.

\input{tables/appendix/reg_robustness/additional_robustness_eval.tex}

\paragraph{System Prompts Robustness}
We evaluate our robustness to system prompts on two leaked system prompts from \textsc{GPT5} and two specially designed adversarial system prompts.
The Robot one corresponds to the Robot personality of \textsc{GPT5} and the second one to the base personality.
We used the system prompts as is but removed the parts related to tooling.
The two additional prompts, Pirate and Weather, are listed below.
\begin{enumerate}
    \item \emph{Pirate}: Talk like a pirate.
    \item \emph{Weather}: Only answer weather-related queries.
\end{enumerate}
\cref{tab:app_results_table} shows that, with realistic system prompts used in production and adversarial system prompts, only our fingerprint is completely robust and maintains a FSR of $1.0$.
For adversarial system prompts in particular, we see that the baselines almost systematically fail, whereas our approach consistently reaches an FSR of 1.0.

\input{figures/appendix/para/para.tex}

\paragraph{Advanced Back-translation}
We further evaluate the robustness of our fingerprint against active adversaries.
Specifically, for both input and output, we consider an advanced back-translation attacker (Translation*) that uses two different models (compared to only one in \cref{sec:eval:robustness}).
In particular, we translate from the prompt language to Chinese using \textsc{GPT5-mini} and then from Chinese back to the prompt language using \textsc{Gemini2.5-Flash-Lite}.
We find in \cref{tab:app_results_table} that our fingerprint consistently reaches an FSR of $1.0$ against such an adversary, unlike both baselines.

\paragraph{Adversarial Paraphrasing}
As shown in \cref{tab:main_results_table}, our fingerprint is robust even against the paraphraser~\citep{diaa2024optimizing} designed to remove the Red-Green watermark, which is the statistical signal used in our fingerprint detection.
In \cref{fig:app_para}, we show the detectability of our fingerprint under paraphrasing and adversarial paraphrasing with respect to the number of queries using \llamabig.
Compared to the baseline deployment, we see that both paraphraser indeed significantly lowers detectability, and that the adversarial paraphraser from~\citet{diaa2024optimizing} is very effective at removing the watermarking signal.
Yet, as the number of queries increases, so does the detectability, and after $\approx 50$ queries we can detect our fingerprint.
This monotonous increase in fingerprint detectability with the number of queries is what enables the strong robustness of our fingerprint.

\subsection{Fingerprinting With Full Watermark}
\label{app:ssec:full_wm_fp}

In this section, we evaluate the advantages of using semantically conditioned watermarks as fingerprints, compared with full (not restricted to a given semantic domain) watermarks, from a robustness perspective.
For stealth, we have already explained in \cref{sec:eval:stealth} the benefits of conditioning on a precise semantic domain.

\input{figures/appendix/full_wm/para.tex}

\paragraph{Setup}
We use a similar setting as in \cref{sec:eval:robustness}.
We embed our fingerprint on \qwen using either French as a semantic domain or by watermarking the model on every domain.
For the latter, we use the same hyperparameters but without the regularization loss, and for the distillation dataset $\mathcal{D}_{\text{target}}$ we use an even mix of the OpenWebText, OpenMathInstruct, and Lucie datasets. 
For detection, we generate up to $1000$ replies using prompts from the French Alpaca dataset.
For finetuning the model, we compare only the non-adversarial case of finetuning outside the semantic domain and perform full finetuning on either Alpaca, OpenMathInstruct, or Dolly. 

\paragraph{Full Watermarking Is Less Robust Against Finetuning}
\cref{fig:app_fullwm} shows that even non-adversarial finetuning (\ie finetuning on a domain different from the one we use for detecting the fingerprint) significantly reduces detectability when using a full watermark instead of a semantically conditioned watermark.\
This means that, surprisingly, in addition to improving the stealthiness of the fingerprint, using semantically conditioned watermarks also contributes to the robustness of the fingerprint compared to full watermarking.

\subsection{Robustness to Instruction-Tuning}
\label{app:ssec:full_wm_fp}

In this section, we evaluate whether (i) we can use our fingerprint on base models and (ii) whether our fingerprint is robust to instruction-tuning (and a subsequent change in prompt format).

\input{figures/appendix/instruction_tuning/instruction_tuning.tex}

\paragraph{Setup}
For embedding our fingerprint (\cref{sec:methods:embedding}) in a base model, we use the same hyperparameters as those described in \cref{app:experimental_details}, but change the semantic domain dataset $\mathcal{D}_{\text{target}}$ and the regularization dataset $\mathcal{D}_{\text{reg}}$ to completion ones.
In particular, we embed our fingerprint in the Health domain and use the health subcategory of the WebOrganizer dataset~\citep{weborganizer} as the target dataset.
For the regularization dataset, we also use the WebOrganizer dataset but without the health subcategory.
As the model, we use the completion version of \qwen instead of the instruction-tuned one.
Then, we instruction-tuned our fingerprinted model using the Qwen chat template on Alpaca.
Specifically, we finetune for one epoch with a batch size of $64$, a learning rate of $2$e$-05$, a warmup of $10$\% with a cosine learning rate scheduler, and the Adafactor optimizer.
To evaluate the robustness of our fingerprint, we then use $1000$ health-related questions extracted from the report in~\citet{health_instruct}, using the newly learnt chat template.
 
\paragraph{Robustness Against Instruction-Tuning}
\cref{fig:app_instruction_tuning} shows that our fingerprint is robust to instruction-tuning and effective on completion models.
Indeed, with $1000$ queries, the FSR is consistently 1.0.
This highlights the inherent robustness of both using a semantic domain as fingerprint queries and a statistical signal as keys.

\subsection{Evaluation of Our Fingerprint Impact on Finetuneability}
\label{app:ssec:full_wm_fp}

In this part, we measure the impact of our fingerprint on finetuneability, \ie whether it is harder to improve performance through finetuning a fingerprinted model compared to the original version.

\input{tables/appendix/reg_finetunability/finetunability_table.tex}

\paragraph{Setup}
We compare our \qwen fingerprinted model on the French domain with the original version.
We independently evaluate finetuneability within the semantic domain and outside the semantic domain.
Within the semantic domain, we finetune both models on WildChatFr for one epoch (using a learning rate of $2$e$-5$, a batch size of $64$, and the Adafactor optimizer with a cosine scheduler and a warmup ratio of $10$\%).
We then measure the benchmark accuracy on FrenchBench before and after finetuning.
Outside the semantic domain, we follow a similar procedure by finetuning both models on AlpacaGalician~\citep{bao2023conversations} (using the same hyperparameters but for $3$ epochs).
We then measure the benchmark accuracy on GalicianBench~\citep{baucells-etal-2025-iberobench} (GB).

\paragraph{Results}
\cref{tab:finetuneability} shows that both without and with our fingerprint, finetuning indeed improves the model performance on the corresponding task.
When finetuning outside the semantic domain (\ie on Galician), we observe that the performance before and after finetuning of the fingerprinted model is the same as that of the original model.
Within the semantic domain (\ie on French), although the performance of the fingerprinted model is lower, it also increases.
Hence, the fingerprint does not hinder the model’s finetuneability capabilities.

%% file: tables/appendix/reg_quality/ppl_and_judge.tex
\begin{table}[t]\centering
    \caption{\textbf{Additional Quality Metrics} We study the impact on quality, measured through perplexity and \textsc{GPT5-mini}-as-a-judge, and report the average across a thousand samples.
    The samples are generated on three different semantic domains: a general Q\&A Domain, on Math questions and on French questions.}
    \label{tab:additional_quality}

    \resizebox{0.8\linewidth}{!}{%
    \begingroup 
    \setlength{\tabcolsep}{5pt}
    \begin{tabular}{ll cc cc cc}
    \toprule
    \multirow{2}{*}{Model} & \multirow{2}{*}{Type} &  \multicolumn{2}{c}{General Q\&A}  &  \multicolumn{2}{c}{Math} &  \multicolumn{2}{c}{French}\\
    \cmidrule(lr){3-4}
    \cmidrule(lr){5-6}
    \cmidrule(lr){7-8}
    && PPL & GPT5 & PPL & GPT5 & PPL & GPT5 \\ 
    \midrule
    \multirow{2}{*}{\llama} & Base & 1.36 & 5.78 & 1.62 & 6.45 & 1.77 & 5.10 \\
    & Fingerprinted & 1.37 & 5.52 & 1.60 & 6.23 & 1.96 & 3.18 \\
    \midrule
    \multirow{2}{*}{\qwen} & Base & 1.45 & 6.62 & 0.92 & 7.46 & 1.72 & 6.47 \\
    & Fingerprinted & 1.50 & 6.52 & 0.91 & 7.54 & 1.78 & 5.12 \\
    \midrule
    \multirow{2}{*}{\llamabig} & Base & 1.42 & 6.86 & 1.40 & 8.51 & 1.45 & 6.90 \\
    & Fingerprinted & 1.45 & 6.64 & 1.44 & 8.48 & 1.75 & 4.82 \\
    \bottomrule
    \end{tabular}
    \endgroup
    }
\end{table}

%% file: tables/appendix/reg_robustness/additional_robustness_eval.tex
\begin{table}[t]\centering
  \caption{\textbf{Complementary Robustness Evaluation} We compare the Fingerprint Success Rate of \llama. \qwen and \llamabig models fingerprinted with either IF, SF or our fingerprint; after various modifications. We highlight in green FSR value of $1.0$, as in \cref{tab:main_results_table}.}
  \label{tab:app_results_table}

  \newcommand{\onecol}[1]{\multicolumn{1}{c}{#1}}
  \newcommand{\twocol}[1]{\multicolumn{2}{c}{#1}}
  \newcommand{\threecol}[1]{\multicolumn{3}{c}{#1}} 
  \newcommand{\fourcol}[1]{\multicolumn{4}{c}{#1}}
  \newcommand{\fivecol}[1]{\multicolumn{5}{c}{#1}}
  \newcommand{\sixcol}[1]{\multicolumn{6}{c}{#1}}
  \newcommand{\sevencol}[1]{\multicolumn{7}{c}{#1}}
  \newcommand{\eightcol}[1]{\multicolumn{8}{c}{#1}}
  \newcommand{\ninecol}[1]{\multicolumn{9}{c}{#1}}

  \renewcommand{\arraystretch}{1.2}
  \newcommand{\skiplen}{0.00001\linewidth} 
  \newcommand{\rlen}{0.01\linewidth} 
  \resizebox{0.8\linewidth}{!}{%
  \begingroup 
  \setlength{\tabcolsep}{5pt} %
  \begin{tabular}{lll ccc p{\skiplen} ccc p{\skiplen} ccc}

      \toprule
      &&& \threecol{\llama} && \threecol{\qwen} && \threecol{\llamabig}\\
    \cmidrule{4-6}
    \cmidrule{8-10}
    \cmidrule{12-14}
     \threecol{\textbf{Modification}}& IF & SF & \textbf{Ours} && IF & SF & \textbf{Ours}&& IF & SF & \textbf{Ours}\\
     \midrule
     \multirow{4}{*}{Sampling} & \multirow{4}{*}{\shortstack{System\\Prompts}} & Robot & \cellcolor{green!25}1.0 & \cellcolor{red!25}0.0 & \cellcolor{green!25}1.0 && \cellcolor{red!25}0.2 & \cellcolor{red!25}0.0 & \cellcolor{green!25}1.0 && \cellcolor{green!25}1.0 & \cellcolor{green!25}1.0 & \cellcolor{green!25}1.0 \\
     && OAI & \cellcolor{green!25}1.0 & \cellcolor{red!25}0.0 & \cellcolor{green!25}1.0 && \cellcolor{red!25}0.8 & \cellcolor{red!25}0.0 & \cellcolor{green!25}1.0 && \cellcolor{green!25}1.0 & \cellcolor{green!25}1.0 & \cellcolor{green!25}1.0 \\
     && Pirate & \cellcolor{red!25}0.0 & \cellcolor{red!25}0.0 & \cellcolor{green!25}1.0 && \cellcolor{red!25}0.0 & \cellcolor{red!25}0.0 & \cellcolor{green!25}1.0 && \cellcolor{red!25}0.0 & \cellcolor{red!25}0.8 & \cellcolor{green!25}1.0 \\
     && Weather & \cellcolor{red!25}0.0 & \cellcolor{red!25}0.0 & \cellcolor{green!25}1.0 && \cellcolor{red!25}0.0 & \cellcolor{red!25}0.0 & \cellcolor{green!25}1.0 && \cellcolor{red!25}0.0 & \cellcolor{green!25}1.0 & \cellcolor{green!25}1.0 \\
     \midrule
     \multirow{2}{*}{Active} & Input & Translation* & \cellcolor{red!25}0.0 & \cellcolor{red!25}0.0 & \cellcolor{green!25}1.0 && \cellcolor{red!25}0.0 & \cellcolor{red!25}0.0 & \cellcolor{green!25}1.0 && \cellcolor{red!25}0.0 & \cellcolor{red!25}0.0 & \cellcolor{green!25}1.0 \\
     & Output & Translation* & \cellcolor{red!25}0.0 & \cellcolor{red!25}0.4 & \cellcolor{green!25}1.0 && \cellcolor{red!25}0.0 & \cellcolor{red!25}0.0 & \cellcolor{green!25}1.0 && \cellcolor{red!25}0.0 & \cellcolor{red!25}0.0 & \cellcolor{green!25}1.0 \\
    \bottomrule 
    \end{tabular}
    
  \endgroup
  }
\end{table}

%% file: figures/appendix/para/para.tex
\begin{figure}[t]
    \centering
    \includegraphics[width=0.85\linewidth]{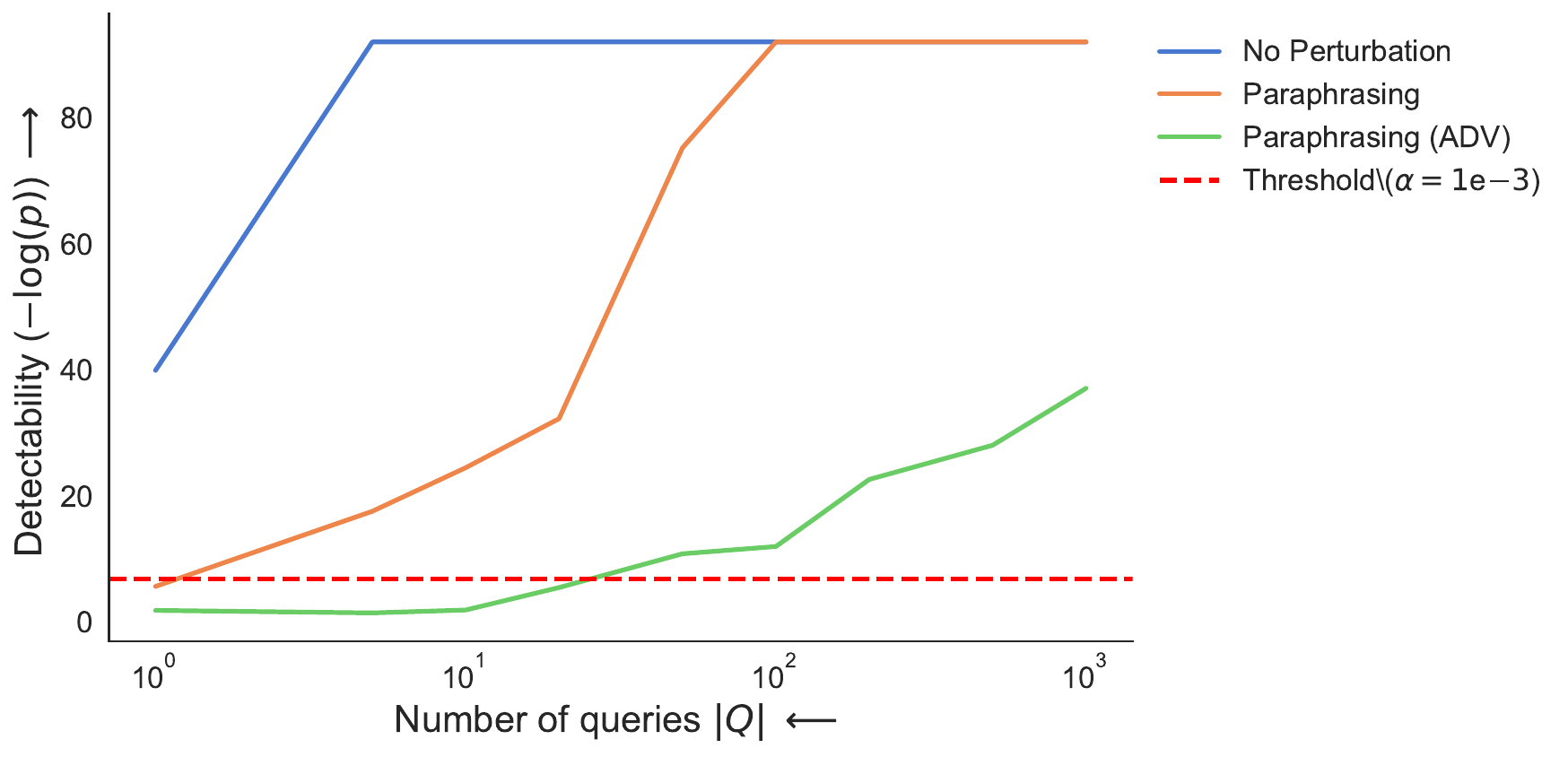}
    \caption{\textbf{Detectability Against Adversarial Paraphrasing} 
    We compare the detectability of our fingerprint (measured by the negative log p-value) with respect to the number of queries $|Q|$ after paraphrasing and adversarial paraphrasing~\citep{diaa2024optimizing}.
    We generate the replies with \llamabig, and average all results over $5$ independent runs.
    In red, we show the fingerprint decision threshold of $1$e$-03$.}
    \label{fig:app_para}
\end{figure}

%% file: figures/appendix/full_wm/para.tex
\begin{figure}[t]
    \centering
    \includegraphics[width=0.85\linewidth]{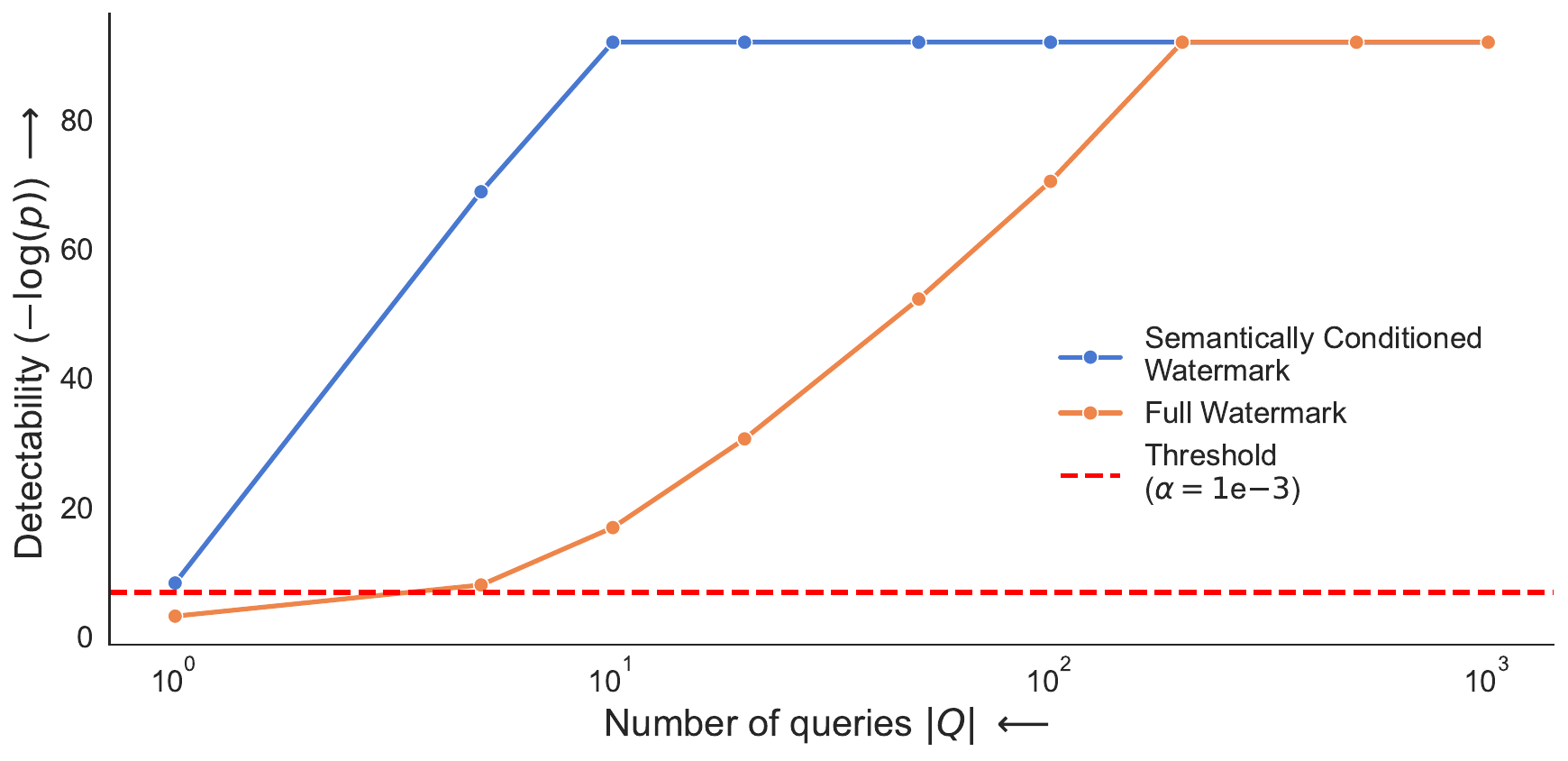}
    \caption{\textbf{Robustness Against Finetuning With/Without Semantically Conditioned Watermarks} 
    We compare the detectability of our fingerprint with semantically conditioned watermarking and with full watermarking (measured by the negative log p-value) with respect to the number of queries $|Q|$ after finetuning~\citep{diaa2024optimizing}.
    We generate the replies with \qwen finetuned on Alpaca, Dolly, or OpenMathInstruct, $5$ times independently and average all results.
    In red, we show the fingerprint decision threshold of $1$e$-03$.
    }
    \label{fig:app_fullwm}
\end{figure}

%% file: figures/appendix/instruction_tuning/instruction_tuning.tex
\begin{figure}[t]
    \centering
    \includegraphics[width=0.8\linewidth]{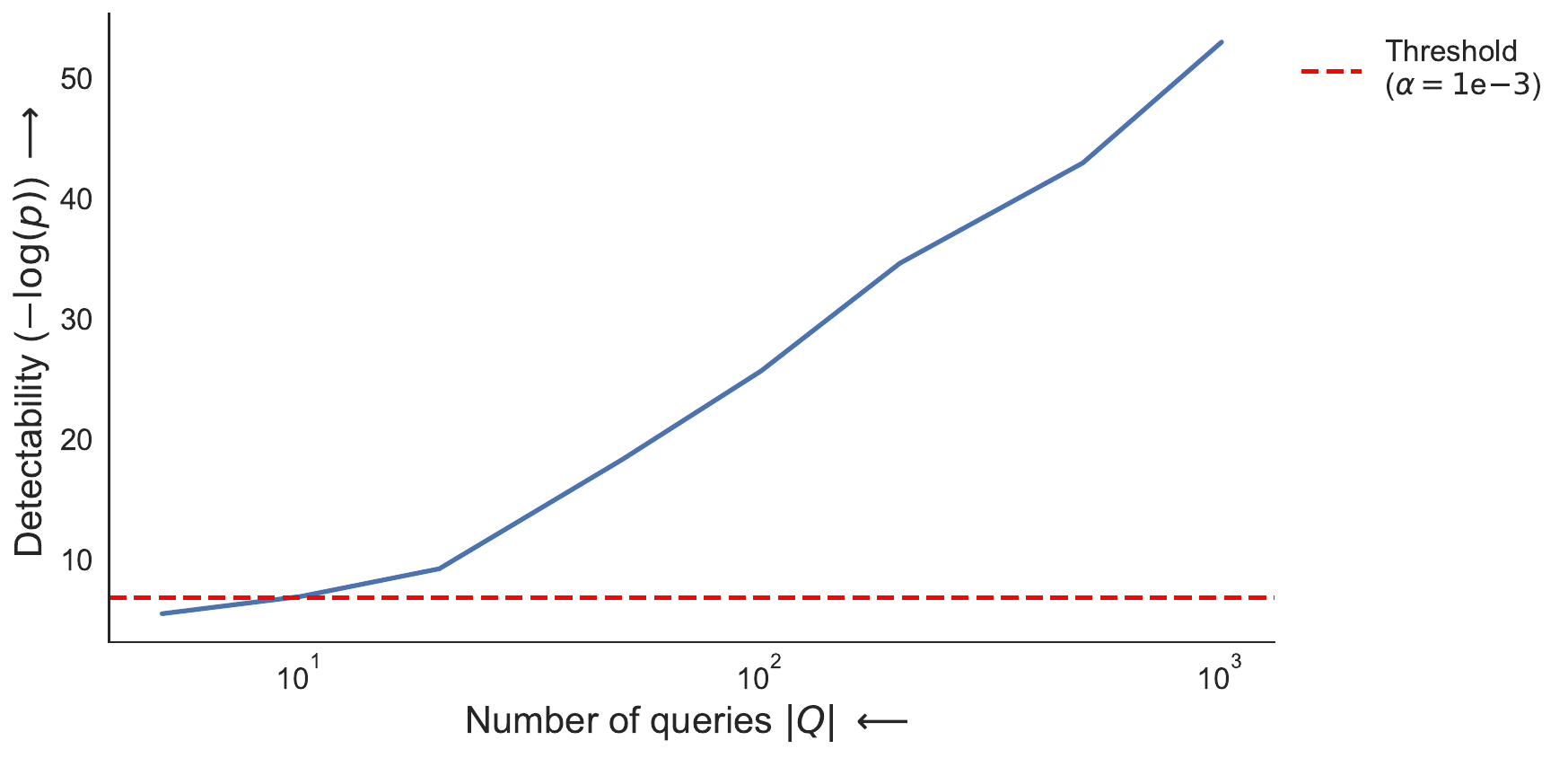}
    \caption{\textbf{Robustness to Instruction-Tuning} 
    We show the detectability of our fingerprint (measured by the negative log p-value) with respect to the number of queries $|Q|$ after instruction-tuning. 
    We generate the replies with \qwen, and average all results over $5$ independent runs.
    In red, we show the fingerprint decision threshold of $1$e$-03$.}
    \label{fig:app_instruction_tuning}
\end{figure}

%% file: tables/appendix/reg_finetunability/finetunability_table.tex
\begin{wraptable}{r}{0.63\textwidth}
\vspace{-1em}
    \caption{\textbf{Evaluation of Our Fingerprint Impact on Finetuneability} We compare the benchmark performance of \qwen models (without/with our fingerprint on French) on task-specific benchmarks (respectively GalicianBench and FrenchBench) before and after finetuning on the corresponding datasets (respectively AlpacaGalician and WildChatFr).}
    \label{tab:finetuneability}
    \resizebox{0.63\textwidth}{!}{%
    \begin{tabular}{lcc||cc}
    \toprule
    \multirow{2}{*}{Model} & \multicolumn{2}{c}{GalicianBench} & \multicolumn{2}{c}{FrenchBench} \\
    \cmidrule{2-3}
    \cmidrule{4-5}
    & Before & After & Before & After \\
    \midrule
    \qwen{} & 46 & 50 & 59 & 63 \\
    \qwen{} (Fingerprinted) & 46 & 51 & 57 & 59 \\
    \bottomrule
    \end{tabular}}
\vspace{-1em}
\end{wraptable}

%% file: src/appendix/app_fingerprint_components.tex
\section{Ablation on Our Fingerprinting Method Components}

We ablate the different component of our method (\cref{sec:methods}), namely the number of queries $|Q|$ required for detection (\cref{app:ssec:monotonicity}), the watermark strength parameter $\delta$ used when embedding the fingerprint (\cref{app:ssec:delta}), the semantic domain on which the fingerprint is embedded (\cref{app:ssec:domain}), the effect of the regularization loss from \cref{eq:dmwm_reg} (\cref{app:ssec:regularization} and \cref{app:ssec:leakage}).

\subsection{Ablation on the Number of Queries}
\label{app:ssec:monotonicity}

In this part, we analyze the dependency between the detectability of our fingerprint, defined as the inverse of the watermark detector p-value, and the number of concatenated queries $Q$ (\cref{sec:methods:detection}).
We find that, in practice, a thousand queries are more than enough to guarantee a robust fingerprint even under adversarial deployment scenarios (\eg finetuning on the semantic domain).

\paragraph{Setup}
We use a similar setting as in \cref{sec:eval:robustness}.
We embed our fingerprint on \llamabig using French as a semantic domain.
For detection, we generate up to $1000$ replies using prompts from the French Alpaca dataset.
We consider two perturbations: finetuning and output paraphrasing.
For the finetuned model, we consider the most adversarial case of finetuning on the semantic domain, and perform full finetuning on the French subset of Wildchat.
For paraphrasing, as in \cref{sec:eval:robustness}, we use \textsc{GPT5-mini} as a paraphraser.

\input{figures/appendix/number_queries/n_queries.tex}

\paragraph{Results}
\cref{fig:app_n_queries} shows that detectability increases with the number of queries.
For all three settings evaluated, we see that the fingerprint is detected with $100$ queries already.
This highlights the strength of using a statistical signal: even in cases where the signal is degraded, we can recover a strong fingerprint signal simply by scaling the number of queries.
Moreover, the base model fingerprint detectability (orange line) is flat, showing that our fingerprint is principled and does not induce an abnormal false positive rate. 

\subsection{Ablation on the Semantic Domain}
\label{app:ssec:domain}

In this part, we show that our fingerprint is effective in additional semantic domains, namely Math and Medicine, and that it also works for completion models (for \cref{sec:evaluation}, we evaluated instruction-tuned models given their prominence).
Additionally, we show that the choice of the semantic domain is crucial for detecting the fingerprint.
For instance, when using LLM watermarks, only domains with high entropy allow strong detectability.

\paragraph{Setup}
For embedding our fingerprint (\cref{sec:methods:embedding}) in the different domains, we use the same hyperparameters and regularization datasets as those described in \cref{app:experimental_details}, but change the semantic domain dataset $\mathcal{D}_{\text{target}}$.
In particular, for the Math fingerprint we use the OpenMathInstruct dataset~\citep{openmathinstruct} as $\mathcal{D}_{\text{target}}$, and for the Medicine fingerprint we use the health subcategory of the WebOrganizer dataset~\citep{weborganizer}.
As the model, we use \qwen and for the Medicine domain we use the completion version of \qwen instead of the instruction-tuned one.
To evaluate the effectiveness of our fingerprint, we follow the same approach as in \cref{sec:eval:effectiveness} and evaluate the fingerprinted models on $8$ LLM benchmarks.
For the Math domain, we use $1000$ prompts from the GSM8K benchmark as $Q$, and for the Medicine domain we use $1000$ extracts of $200$ tokens each from the Medicine wiki~\citep{wikidump}.

\input{tables/appendix/reg_domains/fingerprint_effectiveness.tex}
\input{figures/appendix/domains/domains.tex}

\paragraph{Our Fingerprint is Effective on All Domains}
\cref{tab:app_domain_effectiveness} shows that for all semantic domains we achieve a Fingerprint Success Rate of $1.0$.
Moreover, independently of the domain, we observe no significant drop in benchmark accuracies.
These results indicate that our fingerprint has a minor impact on the model utility, even in more technical domains such as mathematics.

\paragraph{Semantic Domains Should be High Entropy}
\cref{fig:app_domains} shows that the detectability with respect to the number of queries scales differently across domains.
This is because LLM watermark detectability (\ie the detection power of the watermark) is bounded by the entropy of the LLM distribution~\citep{kgw}.
Hence, the entropy of the underlying semantic domain is critical when choosing the domain.
On the positive side, we see that for all evaluated domains increasing the number of queries indeed increases the fingerprint detectability.

\subsection{Ablation on the Watermark Strength Parameter}
\label{app:ssec:delta}

In this part, we analyze the effect of the watermark strength parameter $\delta$ on our fingerprinting method.
We find that a lower $\delta$ reduces detectability (\ie a higher number of queries is needed to recover the fingerprint).
At the same time, a lower $\delta$ also reduces the distortion induced in the model distribution on the semantic domain.
Hence, the choice of $\delta$ allows to balance the cost of detection (because more queries are needed) with the fingerprint impact on model utility.

\input{figures/appendix/delta/delta_figures.tex}

\paragraph{Setup}
We embed our fingerprint in \qwen on the French domain using the same hyperparameters as described in \cref{sec:evaluation}, but with a different watermark strength parameter $\delta$, ranging from $0.5$ to $4$.
To evaluate fingerprint detectability, we query the model up to $1000$ times with prompts from the French Alpaca dataset.
To evaluate the impact of $\delta$ on quality in the semantic domain, we compute the perplexity of the generated replies using \textsc{Qwen2.5-32B}, and the benchmark accuracy on FrenchBench.

\paragraph{Results}
\cref{fig:app_delta} shows that increasing $\delta$ indeed significantly increases detectability, and that $1000$ queries are enough to detect the fingerprint for all $\delta$ tested.
The right side of \cref{fig:app_delta} shows that, at the same time, increasing $\delta$ also slightly increases perplexity in the semantic domain.
Similarly, increasing $\delta$ also slightly decreases FrenchBench accuracy (from $0.61$ at $\delta=0.5$ to $0.57$ at $\delta=5$).
This means that, for the model provider, there is a trade-off between the cost of detectability (due to the increased number of queries needed to detect the watermark) and the impact on quality.
Given that the impact on quality is minimal even with $\delta =4$ (see the benchmark results from \cref{sec:eval:effectiveness}) and limited to the semantic domain, we argue that detectability is more important and thus decided to use $\delta=4$ for our main experiments in \cref{sec:evaluation}

\subsection{Ablation on Our Fingerprint Regularization Loss}
\label{app:ssec:regularization}

In this part, we show that the regularization loss (\cref{eq:dmwm_reg}) is critical for reducing our fingerprint impact on utility and increasing its stealth against targeted adversaries, and also slightly improve robustness against finetuning.

\input{tables/appendix/reg_ablation/fingerprint_effectiveness_reg.tex}

\paragraph{Setup}
We use a similar setup as in \cref{sec:eval:effectiveness,sec:eval:robustness}.
We embed on \qwen our fingerprint using French as a semantic domain but without the regularization loss (\cref{eq:dmwm_reg}).
For detection, we use $|Q|=1000$ different queries from the French Alpaca dataset.

\paragraph{Utility and Stealth}
\cref{tab:reg_fp_effectiveness} shows the effectiveness of our fingerprint with and without regularization.
While the absence of regularization does not alter the detectability of the fingerprint (FSR remains at 1), we notice a slight degradation in model utility despite the watermark distillation loss being applied only on the French dataset.
This already highlights the importance of the regularization term in our fingerprinting method.
Additionally, as mentioned in \cref{sec:eval:stealth}, using a semantic domain is important for stealth as it prevents targeted watermark detection attacks~\citep{detwm,crafted}.

\subsection{Ablation on Fingerprint Leakage}
\label{app:ssec:leakage}

In this part, we show that the fingerprint can only be detected in the semantic domain, and that querying the fingerprinted model on other domains does not leak the fingerprint.
This is important to ensure the stealth benefits of semantically conditioning the fingerprint, detailed in \cref{sec:eval:stealth}.

\input{figures/appendix/leakage/leakage.tex}

\paragraph{Setup}
We use a similar setup as in \cref{app:ssec:domain}.
We embed on \qwen our fingerprint using math as a semantic domain but without the regularization loss (\cref{eq:dmwm_reg}).
For detection, we use $|Q|=1000$ different queries from the GSM8K benchmark (Math).
To test for fingerprint leakage, we also use $|Q|=1000$ different queries from the CodeAlpaca~\citep{codealpaca} dataset (Code), $|Q|=1000$ different health-related queries extracted from~\citet{health_instruct} (Medicine), and $|Q|=1000$ different queries from the Alpaca~\citep{alpaca} dataset (Q\&A).

\paragraph{No Fingerprint Leakage} 
\cref{fig:app_leakage} shows that there is no fingerprint leakage. 
Even when using $1000$ queries for fingerprint detection, only the queries in the semantic domain, \ie math, contain a detectable watermark signal.
This means that, for a targeted adversary to detect the fingerprint (such as in \citet{detwm}), they would indeed need to know the semantic domain beforehand.

%% file: figures/appendix/number_queries/n_queries.tex
\begin{figure}[t]
    \centering
    \includegraphics[width=0.85\linewidth]{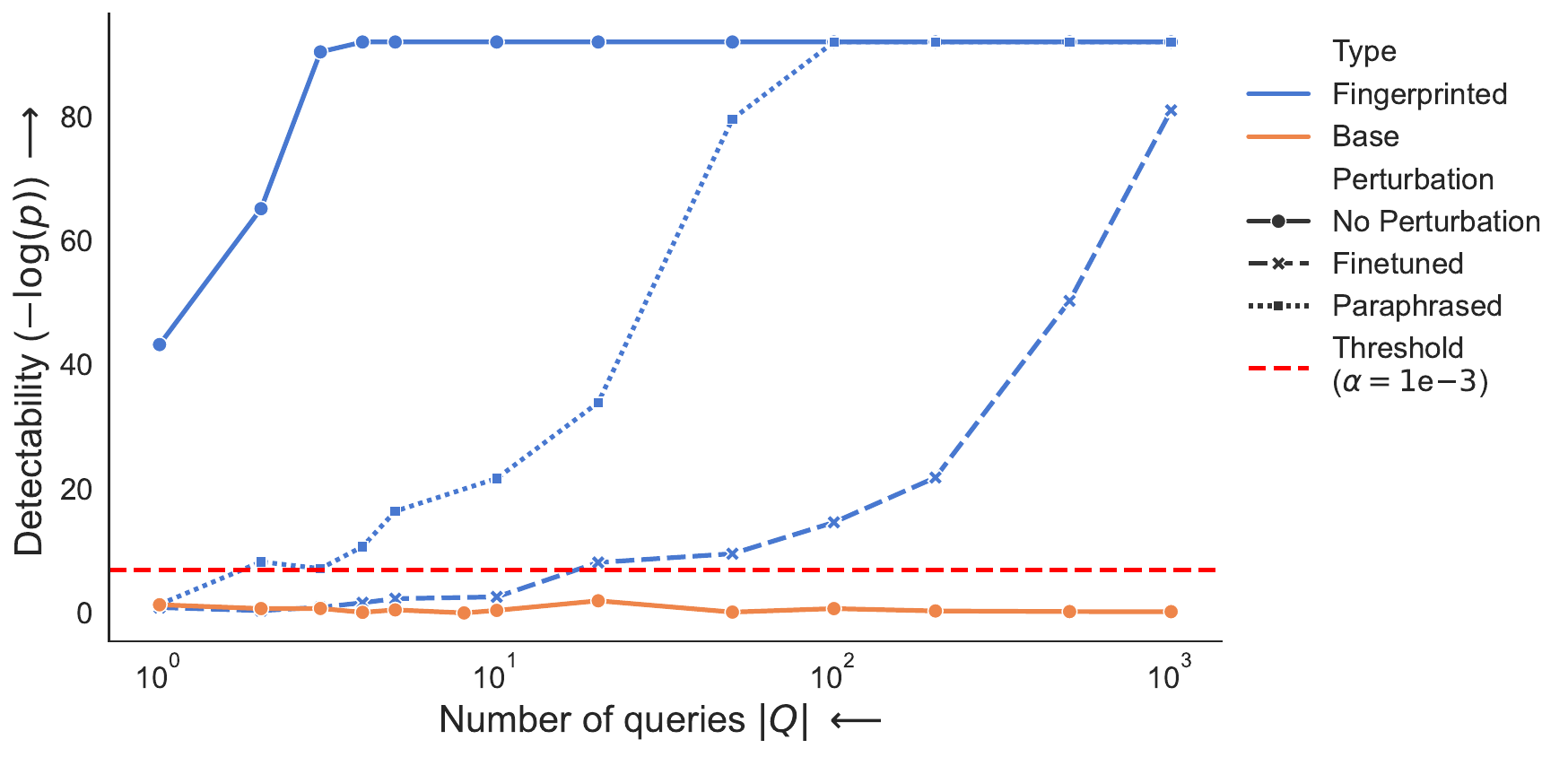}
    \caption{\textbf{Ablation on the Number of Queries} 
    We compare the detectability of our fingerprint (measured by the negative log p-value) with respect to the number of queries $|Q|$.
    We generate the replies with \llamabig, and average all results over $5$ independent runs.
    For the finetuned model, we do full finetuning on the French subset of WildChat and for the paraphrased, we paraphrase the output with \textsc{GPT5-mini}.
    In red, we show the fingerprint decision threshold of $1$e$-03$.}
    \label{fig:app_n_queries}
\end{figure}

%% file: tables/appendix/reg_domains/fingerprint_effectiveness.tex
\begin{table}[t]\centering
    \caption{\textbf{Effectiveness of Our Fingerprinting Method On Different Semantic Domains} We compare the Fingerprint Success Rate (FSR) of \qwen fingerprinted on three different domains. 
    We also compare the utility, measured through benchmark accuracy, and report the average accuracy in the last column (AVG).
    We highlight in bold FSR values of $1.0$.
    }
    \label{tab:app_domain_effectiveness}

    \resizebox{\linewidth}{!}{%
    \begingroup 
    \setlength{\tabcolsep}{5pt}

    \begin{tabular}{ll c cccccccccc}
    \toprule
    \multirow{2}{*}{Model} & \multirow{2}{*}{Domain} &  & \multicolumn{9}{c}{Benchmark Accuracies} \\
    \cmidrule{4-12}
    && \textbf{FSR} & ARC & MMLU & HeSw & TQA & HE & PM-QA & GSM8K & FB & \textbf{AVG}\\ 
    \midrule
    \multirow{4}{*}{\qwen} & Base & 0.0 &0.69 & 0.38 & 0.55 & 0.42 & 0.73 & 0.70 & 0.61 & 0.59 & 0.58 \\
    \cmidrule{2-12}
    & French & \textbf{1.0} & 0.70 & 0.39 & 0.55 & 0.42 & 0.73 & 0.70 & 0.62 & 0.57 & 0.58 \\
    & Medicine & \textbf{1.0} & 0.69 & 0.39 & 0.54 & 0.42 & 0.76 & 0.69 & 0.61 & 0.59 & 0.59 \\
    & Math & \textbf{1.0} & 0.70 & 0.38 & 0.54 & 0.43 & 0.74 & 0.68 & 0.56 & 0.59 & 0.58 \\
    \bottomrule
    \end{tabular}
    \endgroup
    }
\end{table}

%% file: figures/appendix/domains/domains.tex
\begin{figure}[t]
    \centering
    \includegraphics[width=0.85\linewidth]{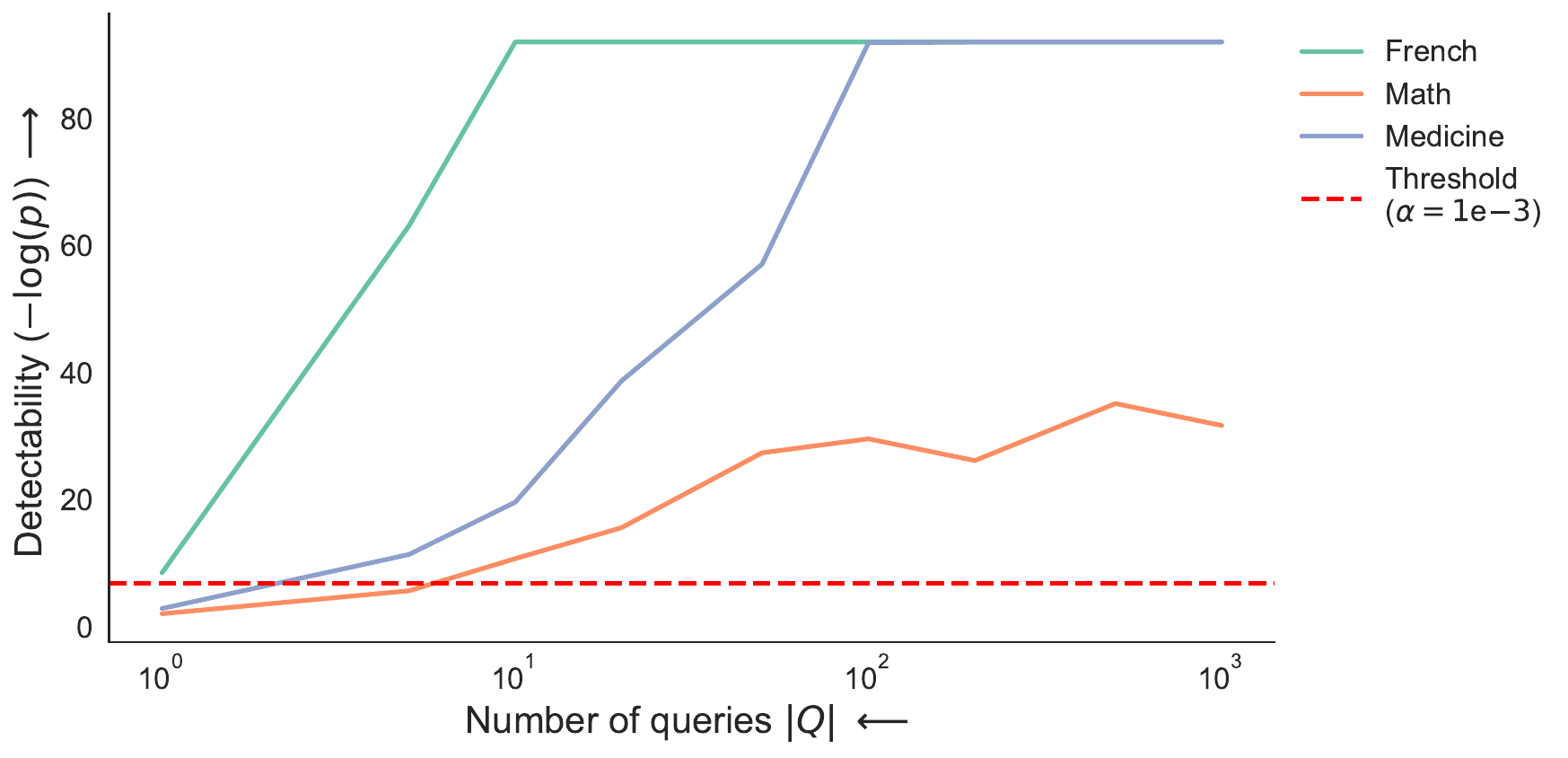}
    \caption{\textbf{Ablation on the Semantic Domain} 
    We compare the detectability of our fingerprint (measured by the negative log p-value) with respect to the number of queries $|Q|$ for different semantic domains.
    We generate the replies with \qwen, and average all results over $5$ independent runs.
    In red, we show the fingerprint decision threshold of $1$e$-03$.}
    \label{fig:app_domains}
\end{figure}

%% file: figures/appendix/delta/delta_figures.tex
\begin{figure}[t]
    \centering
    \includegraphics[width=0.60\linewidth]{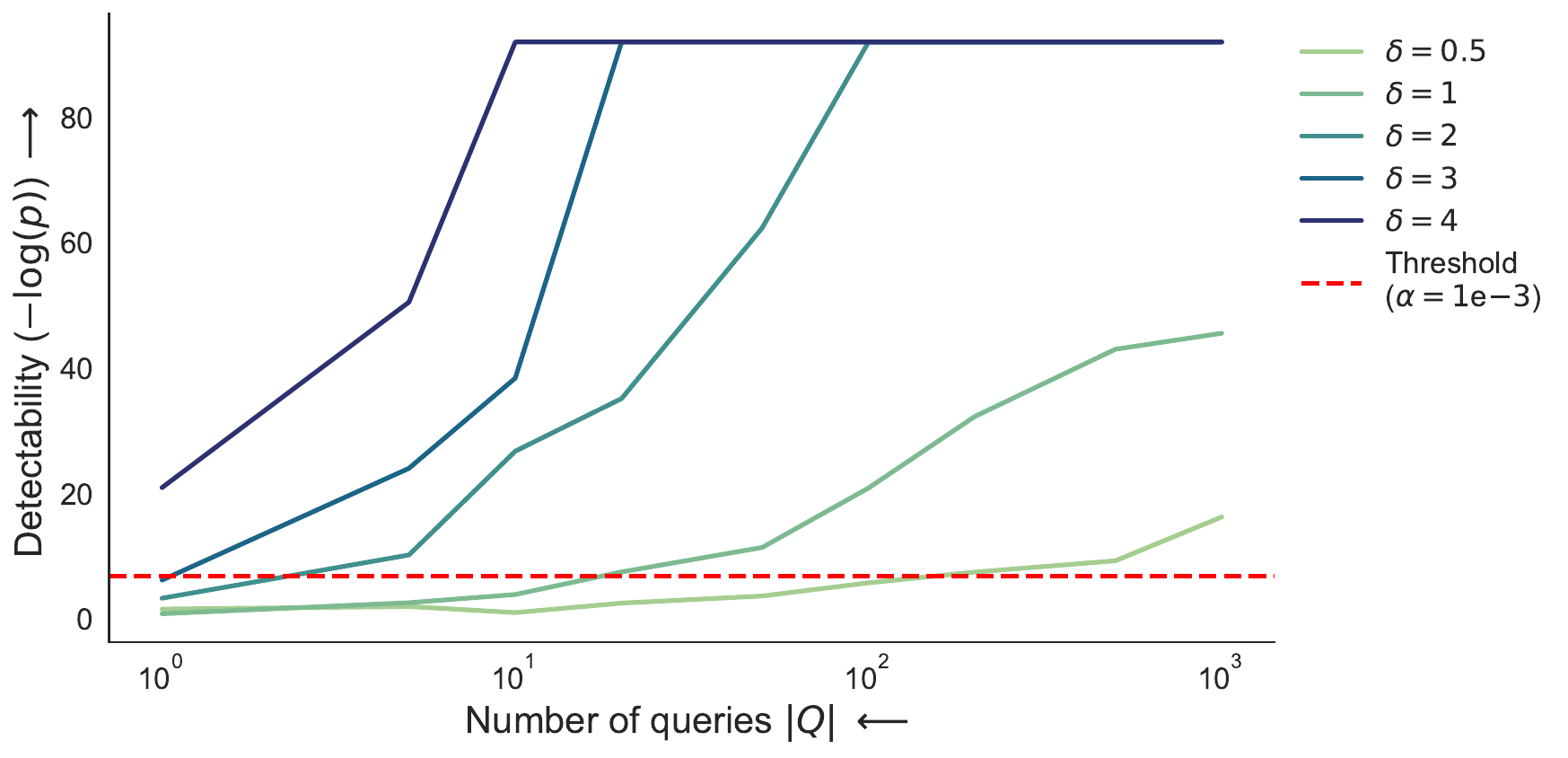}
    \includegraphics[width=0.38\linewidth]{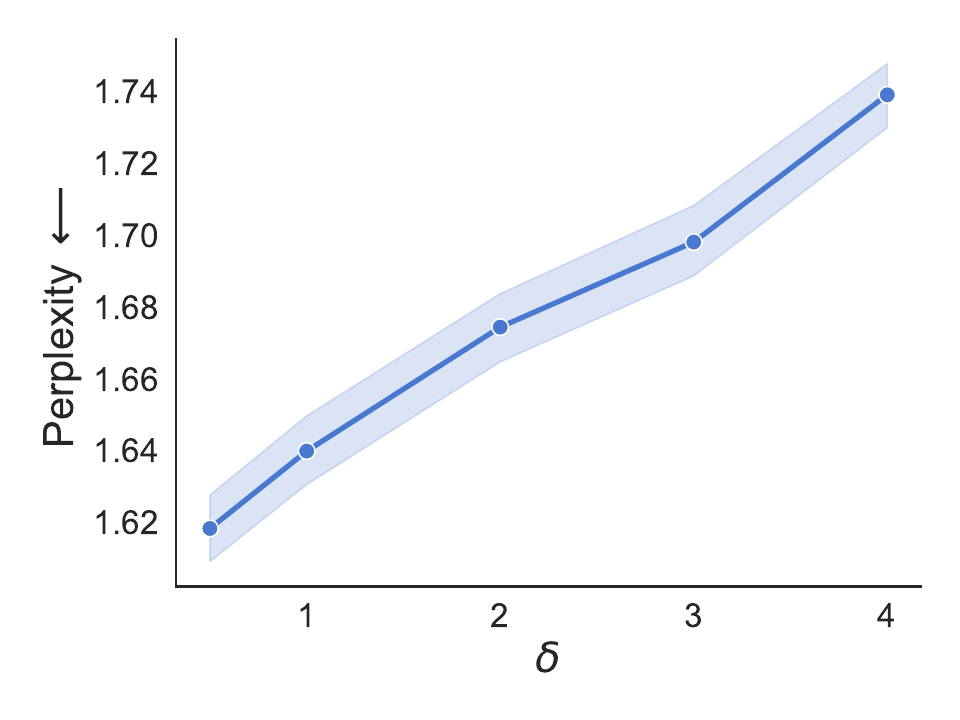}
    \caption{\textbf{Ablation on the Watermark Strength Parameter} (\emph{Left}) We compare the detectability of our fingerprint (measured by the negative log p-value) with respect to the number of queries $|Q|$ for different watermark strength parameters $\delta$ (\cref{sec:methods:embedding}). 
    (\emph{Right}) We study the dependency between our fingerprint's impact on quality (measured through PPL computed with \textsc{Qwen2.5-32B}) and the watermark strength parameter $\delta$.
    For both figures, we generate $1000$ replies with \qwen fingerprinted on the French domain, and average the metrics over $5$ independent runs.}
    \label{fig:app_delta}
    \vspace{-0.15in}
\end{figure}

%% file: tables/appendix/reg_ablation/fingerprint_effectiveness_reg.tex
\begin{table}[t]\centering
    \caption{\textbf{Effectiveness of Our Fingerprinting Method} We compare the Fingerprint Success Rate (FSR) of our method with (Ours) and without the regularization (Ours (w/o)).
    We also compare the utility, measured through benchmark accuracy, and report the average accuracy in the last column (AVG).
    We highlight in bold FSR values above 80\%.}
    \label{tab:reg_fp_effectiveness}

    \resizebox{\linewidth}{!}{%
    \begingroup 
    \setlength{\tabcolsep}{5pt}

    \begin{tabular}{ll c ccccccccc}
    \toprule
    \multirow{2}{*}{Model} & \multirow{2}{*}{Type} &  & \multicolumn{9}{c}{Benchmark Accuracies} \\
    \cmidrule{4-12}
    && \textbf{FSR} & ARC & MMLU & HeSw & TQA & HE & PM-QA & GSM8K & FB & \textbf{AVG}\\ 
    \midrule
    \multirow{3}{*}{\qwen} & Base & 0.0 &0.69 & 0.38 & 0.55 & 0.42 & 0.73 & 0.70 & 0.61 & 0.59 & 0.58  \\
    & Fingerprinted & \textbf{1.0} & 0.70 & 0.39 & 0.55 & 0.42 & 0.73 & 0.70 & 0.62 & 0.57 & 0.58 \\
    & Ours (w/o) & \textbf{1.0} & 0.70 & 0.38 & 0.54 & 0.40 & 0.66 & 0.70 & 0.61 & 0.57 & 0.57 \\
    \bottomrule
    \end{tabular}
    \endgroup
    }
\end{table}

%% file: figures/appendix/leakage/leakage.tex
\begin{figure}[t]
    \centering
    \includegraphics[width=0.85\linewidth]{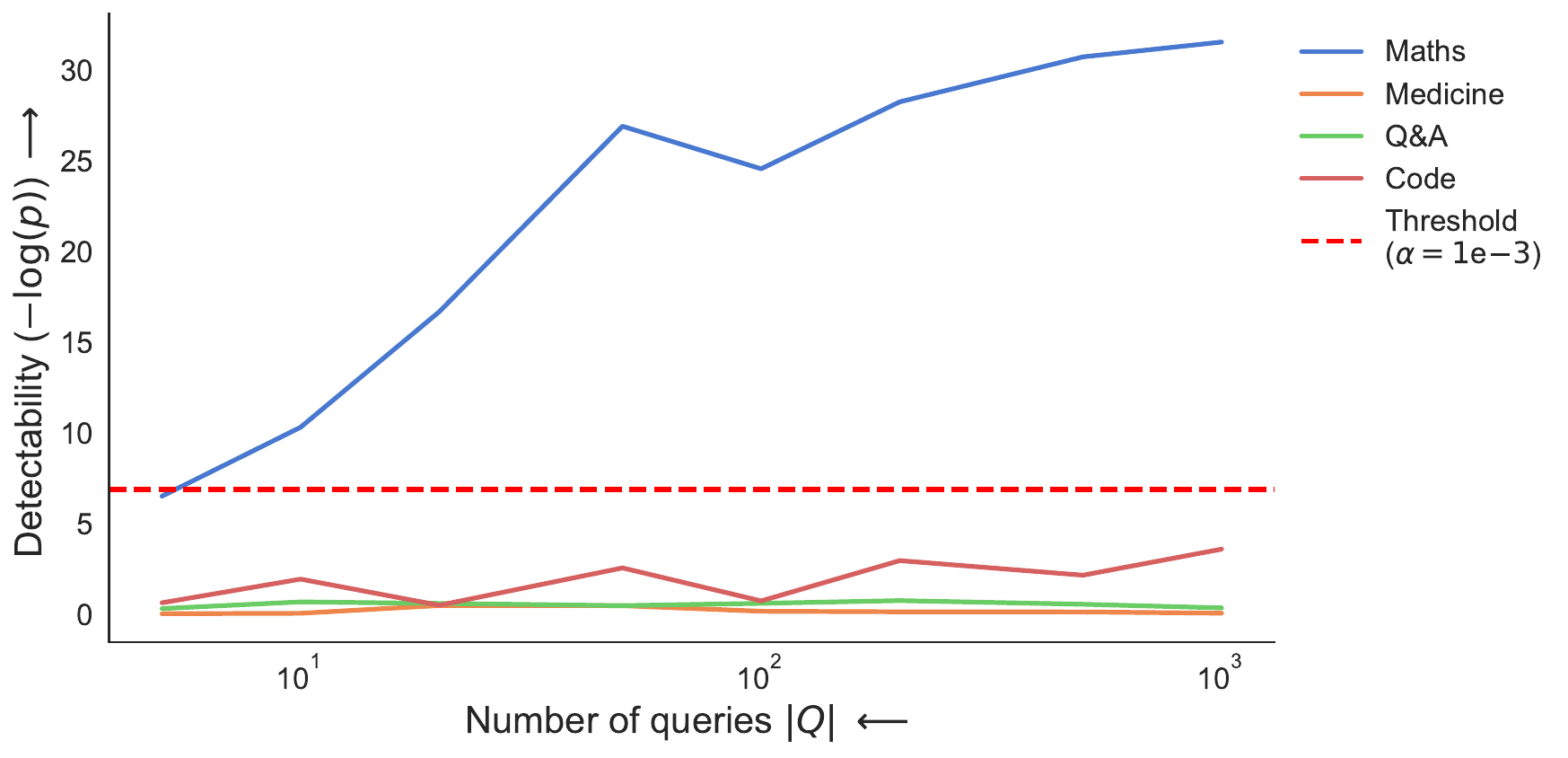}
    \caption{\textbf{Ablation on Fingerprint Leakage} 
    We compare the detectability of our fingerprint (measured by the negative log p-value) with respect to the number of queries $|Q|$ for different semantic domains.
    We generate the replies with \qwen fingerprinted on the math domain, and average all results over $5$ independent runs.
    In red, we show the fingerprint decision threshold of $1$e$-03$.
    We see that only with the math domain is the fingerprint detected: there is no leakage.
   }
    \label{fig:app_leakage}
\end{figure}

%% file: src/appendix/app_fingerprint_detection.tex
\section{Fingerprint Detection}
\label{app:fingerprint_detection}

In this part, we expand on the derivation from \cref{sec:methods:detection} to show that our detector is principled, \ie that the false positive rate is properly controlled.

\paragraph{Red-Green Watermarking}
We recall the notation and (partial) definition of Red-Green watermark from \cref{sec:background_and_related}.
At each step $t$ of the generation process, using both a private key $\xi$ and the previous $k$ tokens, the vocabulary $\Sigma$ is pseudorandomly partitioned into $\gamma |\Sigma|$ green tokens and the rest are red tokens.
We introduce the green list matrix $G \in \{0,1\}^{\Sigma^k \times \Sigma}$ that, for each previous $k$ tokens (and the private key), associates the green list partition ($1$ for green tokens and $0$ for red ones).
With this parametrization, choosing a key $\xi$ is equivalent to sampling $G$ from the following distribution: each row is independent, and within each row the cells are correlated Bernoulli random variables, where the correlation corresponds to enforcing the size of the green list.
Hence we consider $G$ as a random matrix distributed according to
\begin{align}
    &\forall h \in \Sigma^k, \forall t \in \Sigma, G_{h,t} \sim \mathcal{B}(\gamma) \\
    &\forall h \in \Sigma^k, \forall t\neq t' \in \Sigma, \text{Cov}(G_{h,t},G_{h,t'}) = - \frac{\gamma (1 - \gamma)}{|\Sigma| - 1}\\
    &\forall h\neq h'\in \Sigma^k, \forall t, t' \in \Sigma, G_{h,t} \perp G_{h,t'},
\end{align}
where $\perp$ means statistical independence.

\paragraph{Fingerprint Detection}
Let $\omega \in \Sigma^*$ be a sequence of tokens (or a concatenation of sequences of tokens).
From this sequence, we can derive a sequence of tuples $(t_1,h_1),\ldots,(t_n,h_n)$ from $\Sigma\times\Sigma^k$.
We then deduplicate the sequence of pairs $(t_i,h_i)$.
Hence, let $n$ be the number of distinct pairs $(t_i,h_i)$.
Without loss of generality, we assume that the above sequence has no duplicates.
The Z-score from \cref{eq:zscore} can be reformulated as
\begin{equation}
    Z(\omega) = \frac{1}{\beta(\omega)\sqrt{n \gamma (1-\gamma)}} \left(\sum_{i=1}^n G_{h_i,t_i}  - n\gamma\right),
\end{equation}
where $\beta(\omega)$ is the correction term mentioned in \cref{sec:methods:detection}.

\begin{theorem} \label{theorem:asymptot_normality}
    Let $\omega \in \Sigma^*$ a (deduplicated) token sequence sampled independently from $G$. For all $h \in \Sigma^k$, let $m_h := |\{i \le n: h_i = h\}|$. We introduce the effective length
    \begin{equation}
        n_{\mathrm{eff}}(\omega) := \sum_{h \in \Sigma^k} m_h\,\frac{|\Sigma|-m_h}{|\Sigma|-1}.
    \end{equation}  
    Assume that $\max_h m_h = o(\sqrt{n_{\mathrm{eff}}(\omega)})$ and set
    \begin{equation} \label{eq:beta}
        \beta(\omega) := \sqrt{\frac{n_{\mathrm{eff}}(\omega)}{n}},
    \end{equation}
    we have that $Z(\omega)$ follows asymptotically a standard normal distribution.
\end{theorem}

This theorem thus ensures that performing a Z-test on $Z(\omega)$ is principled: the false positive rate is correctly controlled.
This result is what ensures that the Red-Green watermark detector from~\citet{kgw} is principled, though they omit the correction term $\beta$, which makes their test more conservative.
The assumption on $m_h$ simply states that there should be enough context diversity.
We argue this assumption is not restrictive and is likely to be met in natural text.

\begin{proof}
    The strategy of the proof is to show that, with such $\beta(\omega)$, $Z(\omega)$ has mean zero and variance one. 
    Then its asymptotic standard normality follows from the Lindeberg Central Limit Theorem.

    Let $X_i = G_{h_i,t_i}$ and $Y_h = \sum_{i: h_i =h}X_i$ be the sum of green tokens with context $h$.
    We note that, conditioned on $\omega$, the $Y_h$ are independent and follow a hypergeometric distribution with parameters $(|\Sigma|,\gamma|\Sigma|,m_h)$.
    In particular,
    \begin{align}
        &\mathbb{E}[Y_h] = m_h \gamma, \\
        &\text{Var}(Y_h) = m_h \gamma (1-\gamma) \frac{|\Sigma| - m_h}{|\Sigma| - 1}.
    \end{align}

    Let $S_n := \sum_{i=1}^n X_i = \sum_{h \in \Sigma^k} \sum_{i: h_i = h} X_i = \sum_{h \in \Sigma^k} Y_h$.
    By independence of the $Y_h$,
    \begin{align}
        &\mathbb{E}[S_n] = n \gamma, \\
        &\text{Var}(S_n) = \gamma (1-\gamma) n_{\mathrm{eff}}(\omega).
    \end{align}
    Thus we have that $Z(\omega)$ with the $\beta(\omega)$ from \cref{eq:beta} is centered and has variance one.

    Let $H=\{h\in\Sigma^k:m_h>0\}$. We decompose $S_n-\mathbb{E}[S_n]$,
    \begin{equation}    
        S_n-\mathbb{E}[S_n] \;=\; \sum_{h\in H}\bigl(Y_h-\mathbb{E}[Y_h]\bigr)
        \;=:\; \sum_{h\in H} W_h.
    \end{equation}
    For all $h\in\Sigma^k$, let $\sigma_h^2 = \text{Var}(Y_h) = \text{Var}(W_h)$, and let $s_n^2:=\sum_{h\in H}\sigma_h^2=\gamma(1-\gamma)\,n_{\mathrm{eff}}(\omega)$.
    Our assumption $\max_h m_h=o\!\left(\sqrt{n_{\mathrm{eff}}(\omega)}\right)$ implies that $s_n\to\infty$.

    We verify Lindeberg’s condition for the triangular array $\{W_h\}_{h\in H}$:
    for every $\varepsilon>0$,
    \begin{equation}  
        \frac{1}{s_n^2}\sum_{h\in H}
        \mathbb{E}\!\left[\,W_h^{2}\,\mathbf{1}\!\left\{|W_h|>\varepsilon s_n\right\}\right]
        \;\le\;
        \frac{1}{s_n^2}\sum_{h\in H}
        m_h^{2}\,\mathbf{1}\!\left\{m_h>\varepsilon s_n\right\}.
    \end{equation}
    The inequality uses the bound $|W_h|=|Y_h-\mathbb{E}[Y_h]|\le m_h$ since $Y_h\in\{0,\dots,m_h\}$.
    By the assumption $\max_h m_h=o(s_n)$, for any fixed $\varepsilon>0$ and all $n$ large enough we have
    $m_h\le \varepsilon s_n$ for every $h\in H$, hence the indicators vanish and the last sum equals $0$.
    Therefore Lindeberg’s condition holds. Hence, by the Lindeberg–Feller Central Limit Theorem,
    \begin{equation}
        \frac{S_n-\mathbb{E}[S_n]}{s_n} \xrightarrow{d}\mathcal{N}(0,1).
    \end{equation}
    Since $s_n=\sqrt{\gamma(1-\gamma)\,n_{\mathrm{eff}}(\omega)}$, this is exactly $Z(\omega)$ with the choice of $\beta(\omega)$ in \cref{eq:beta}.
    Hence $Z(\omega)\xrightarrow{d} \mathcal{N}(0,1)$, which proves the theorem.
\end{proof}

%% file: src/appendix/app_semantic_watermarks.tex
\section{On Semantically Conditioned Watermarks}
\label{app:semantic_wm}

In this section, we explore the boundaries of semantically conditioned watermarks.
In \cref{app:ssec:wm_token}, we introduce the concept of a watermark token and show that models can associate a domain with the presence of the token, and can even adapt to the setting of an opening and closing watermark token. 
We expand on watermark tokens in \cref{app:ssec:multiple_keys} and show how we can train a model to associate a different key with multiple watermark tokens.
Lastly, in \cref{app:ssec:other_domains}, we propose watermarking the harmful semantic domain.

\paragraph{Setup}
In all subsequent experiments, we use the same optimization hyperparameters as in \cref{sec:evaluation} to train the semantically conditioned watermark on \textsc{Llama-3.2-1B}.
For training hyperparameters, we set the batch size to 64 with 512-token-long sequences, the learning rate to 2e-5 with a cosine scheduler and a 250-step warmup, and we use the Adafactor \citep{adafactor} optimizer.
For Red-Green watermarks, we set $\gamma=0.25$, $\delta = 4.0$, and $k=1$.

\subsection{Watermark Tokens as Domain Trigger}
\label{app:ssec:wm_token}

\input{figures/appendix/semantic_watermark/roc_wm_tokens.tex}

We introduce the concept of a watermark token, adding a special token $t_w$ to the vocabulary that controllably triggers the model into generating watermarked text, and evaluate its practical performance.

\paragraph{Setup}
We use \textsc{OpenWebText} as a training dataset, where we prepend all token sequences with the watermark token $t_w$.
For the regularization dataset, we also use \textsc{OpenWebText} but without the watermark token.
To evaluate the watermark on the watermark token domain, we generate a thousand 200-token-long completions using 50-token-long completion prompts from the RealNewsLike split of the \textsc{C4} dataset \citep{c4}, both with and without $t_w$, following the setup from prior work \citep{kgw}, and compute the watermark p-value with a one-sided Z-test (\cref{eq:zscore}). 

\paragraph{Watermark Token}
In \cref{fig:wm_tokens} (left), we plot the ROC curve of this experiment, and the identity line in gray as a reference.  
We see that the model almost perfectly learns the trigger, outputting a strong watermark when $t_w$ is present (above 95\% TPR at 1\%), while outputting almost no watermarked text in the absence of $t_w$.  
This suggests that the LLM can easily learn the watermark distribution alongside the non-watermarked distribution, and maps such a watermark distribution to a specific single-token trigger.  
While in \cref{sec:evaluation} we only consider domains as triggers, one could use any token combination as a trigger, improving the harmlessness of the fingerprint at the cost of stealthiness, while benefiting from the strong reliability and guarantees of watermarking-based fingerprints.

\paragraph{Opening and Closing Watermark Tokens}
We further expand the notion of a watermark token by adding opening and closing watermark tokens, namely \emph{<wm>} and \emph{</wm>}.
For training, we use \textsc{OpenWebText} as the training dataset, where, for each sample in the dataset, we uniformly sample a contiguous sequence of between 200 and 400 tokens.
We then enclose such sequences in watermark tokens.
Within the enclosed tokens, we use watermark distillation as described in \cref{eq:wm_learning}, and outside the tokens, we use the regularization loss from \cref{eq:dmwm_reg}.
Additionally, we also apply regularization on \textsc{OpenWebText} without any watermark tokens.
To evaluate the success of the jailbrokenwatermarks, we generate a thousand 200-token-long completions using 50-token-long completion prompts from four different variants of the RealNewsLike split of the \textsc{C4} dataset.
We use the raw data, prompts ending with \emph{<wm>}, prompts where we append a watermarked completion enclosed in <wm>…</wm> to the C4 prompt, and similarly for the non-watermarked case.
The last two variations allow us to differentiate between the model actually generating watermarked text and watermark radioactivity \citep{radioactivity}.
We show the pattern used for the different prompts:

\begin{description}[%
    style=nextline,
    labelwidth=\widthof{\bfseries C4 closed (noWM):},
    leftmargin=!]
    \item[\textbf{C4:}]
    \texttt{C4 text + \textcolor{red!50}{starts generating here}}
    \item[\textbf{C4 open:}]
    \texttt{C4 text + <wm> + \textcolor{red!50}{starts generating here}}
  \item[\textbf{C4 closed:}]
    \texttt{C4 text + <wm> watermarked text </wm> + \textcolor{red!50}{starts generating here}}
  \item[\textbf{C4 closed (noWM):}]
    \texttt{C4 text + <wm> normal text </wm> + \textcolor{red!50}{starts generating here}}
\end{description}

In \cref{fig:wm_tokens} (right), we plot the ROC curves across all datasets, as well as the identity line for reference.  
We see that in the absence of watermark tokens (\textcolor{blue}{blue line}), the text generated by the model is non-watermarked.  
When the watermark token is opened (\textcolor{red}{red line}), the model generates watermarked text.  
Lastly, when the watermark token is closed and watermarked text is enclosed (\textcolor{orange}{orange line}), we observe some watermark radioactivity—the model still generates slightly watermarked text.  
In contrast, when the enclosed text is not watermarked (\textcolor{green!30}{green line}), the model does not generate watermarked text, as intended.  
These results show how flexible semantically conditioned watermarking can be.  
Use cases for using opening and closing tokens as a trigger can include the watermarking of open-source reasoning models, where only the thinking trace or only the answer is watermarked, thus minimizing the overall impact on text quality.  
We leave this direction for future work.

\subsection{Watermarking Multiple Token-Conditioned Domains}
\label{app:ssec:multiple_keys}

\input{figures/appendix/semantic_watermark/roc_curves_multiwm.tex}

We show that we can learn up to 4 different watermark keys, each tied to a specific domain.

\paragraph{Experimental setup}
We only consider using watermark tokens as different domains.  
For each watermark token (represented by [WM<i>] where $i\in\{0,1,2,3\}$), we train a different watermarking key, effectively learning a completely new watermark distribution.  
To do so, we proceed as in \cref{app:ssec:wm_token}, where we use up to 4 \textsc{OpenWebText} datasets, each prepended with the corresponding watermark token, and apply logit distillation with the corresponding watermark key, along with an additional \textsc{OpenWebText} dataset without any watermark token as a regularizer.  
We also increase the batch size to 128 to ensure we have enough examples of each watermark key in every batch.  
To evaluate each key, we generate a thousand 200-token-long completions using 50-token-long completion prompts from the RealNewsLike split of the \textsc{C4} dataset, where we prepend the watermarking key.  
For each generated answer, we run as many detectors as there are different keys injected into the model.

\paragraph{Semantically Conditioned Watermarks Generalize to Multiple Keys}
In \cref{fig:wm_multiple_key}, we show the ROC curves, running the detection with every key for each scenario.  
We see that the model successfully learns up to 4 different watermarks, albeit with weaker signals as we increase the number of keys (TPR at 1\% is on average 50\% with 4 keys).  
Moreover, we see that even with the detector corresponding to another key (for instance, detector 1 with [WM2] instead of [WM0]), the detection rate is sometimes abnormally high.  
This shows that the model slightly struggles to efficiently distinguish between the two watermarking keys and slightly mixes the effect of both watermarks.
Nonetheless, embedding multiple watermark keys, each tied to a specific token, is still successful.

\subsection{Watermarking Other Domains (Harmful Content)}
\label{app:ssec:other_domains}

\input{figures/appendix/semantic_watermark/harmful_roc.tex}

In this part, we explore a new semantic domain to watermark: harmful content.  
By watermarking such a domain, we can easily trace harmful text generated by our model in the wild—expanding on prior work in this direction \citep{harmfultoken}.

\paragraph{Setup}
We train an \textsc{Llama3.2-1B} with similar hyperparameters as in \cref{sec:evaluation} using the harmful data from \textsc{LLM-LAT} \citep{llmlat}.
To evaluate the watermark, because the model is aligned, we use prefilling jailbreak~\citep{prefill_jailbreak} with 300 harmful queries, and generate three hundred 200-token-long completions.

\paragraph{Harmful Domain Can Be Watermarked}
In \cref{fig:harmful_roc}, we show the ROC curves for the harmful domain and a general domain (Alpaca).  
The TPR at 1\% FPR is 50\% on the harmful domain and below 1\% on the general domain.  
This means that even when jailbroken, the model still outputs watermarked text.  
These results align with previous work on self-identification of harmful data \citep{harmfultoken}, where a model is trained to add a detectable signal when and only when generating harmful content.

%% file: figures/appendix/semantic_watermark/roc_wm_tokens.tex
\begin{figure}[t]
    \centering
    \includegraphics[width=0.49\textwidth]{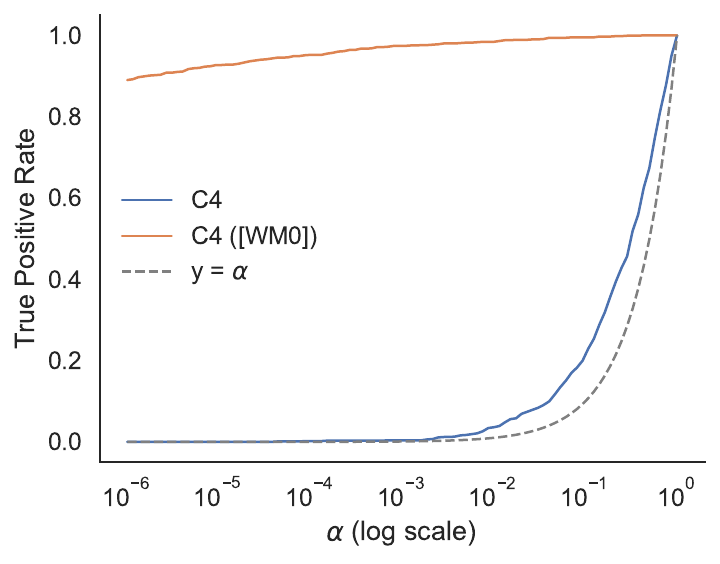}
    \includegraphics[width=0.49\textwidth]{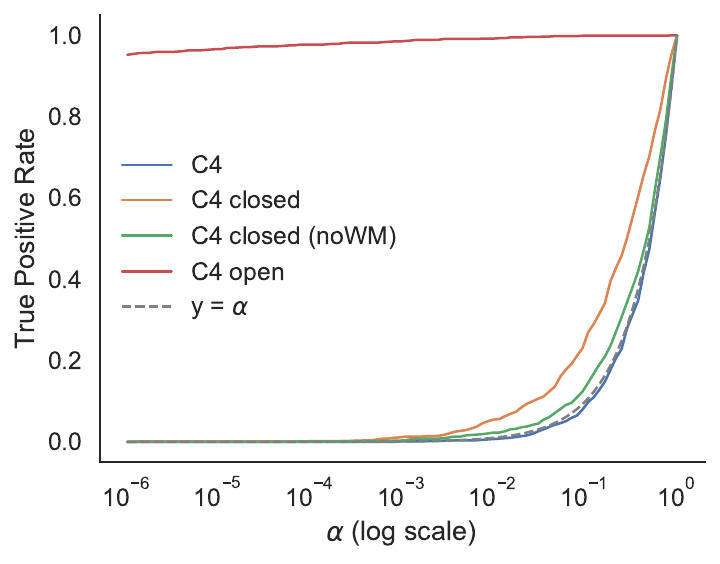}
    \caption{ROC curves for evaluating semantically conditioned watermarks with a watermark token (left) or an (opening,closing) watermark token (right).}
    \label{fig:wm_tokens}
    \vspace{-0.1in}
\end{figure}

%% file: figures/appendix/semantic_watermark/roc_curves_multiwm.tex
\begin{figure}[t]
    \centering
    \includegraphics[width=0.9\textwidth]{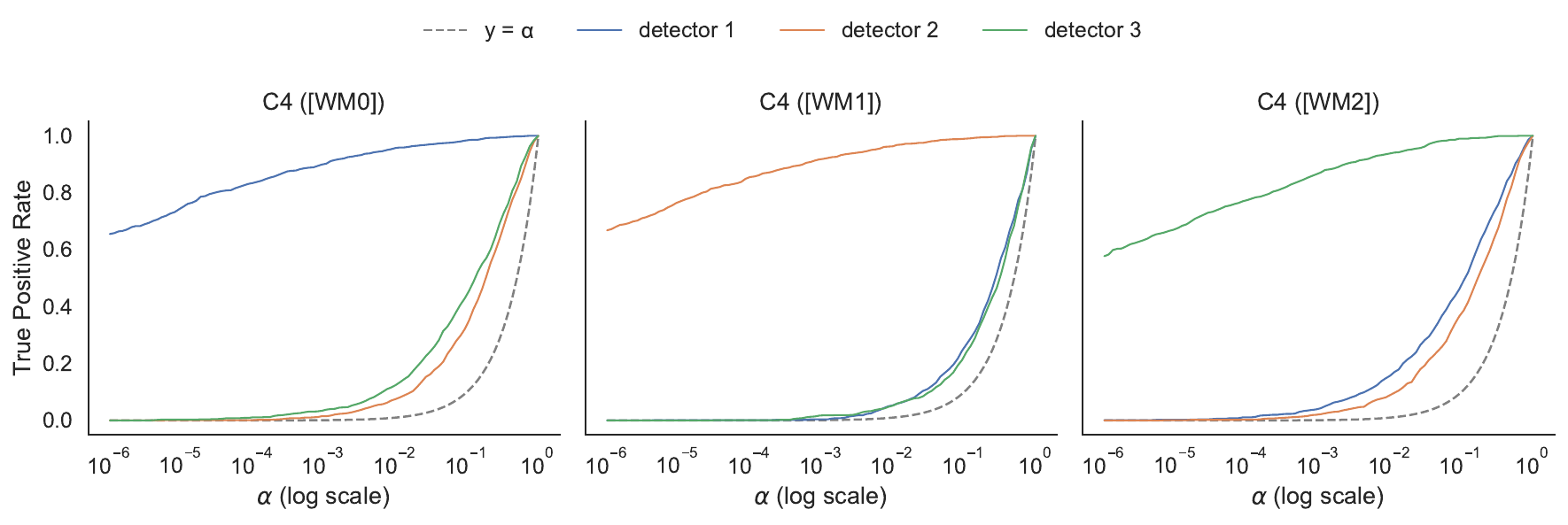}
    \includegraphics[width=\textwidth]{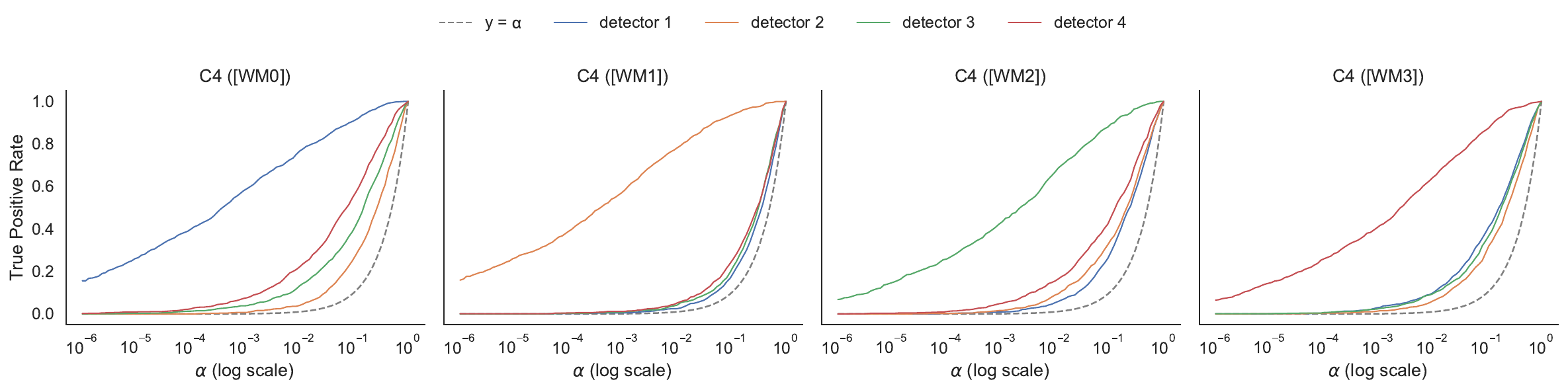}
    \caption{ROC curves for evaluating semantically conditioned watermark with a different key per watermark token.}
    \label{fig:wm_multiple_key}
    \vspace{-0.1in}
\end{figure}

%% file: figures/appendix/semantic_watermark/harmful_roc.tex
\begin{figure}[t]
    \centering
    \includegraphics[width=0.4\textwidth]{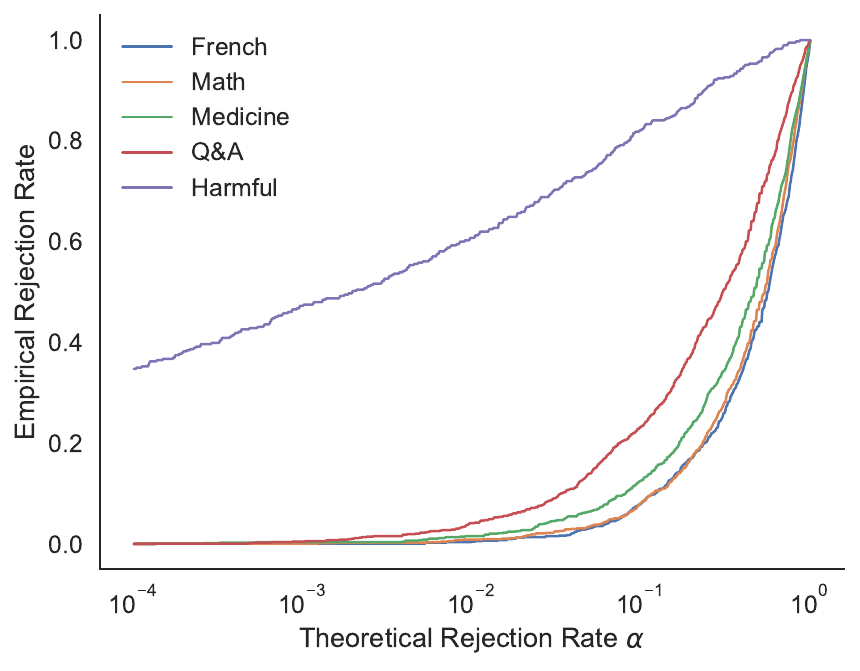}
    \caption{ROC curve for evaluating semantically conditioned watermark on the harmful domain.}
    \label{fig:harmful_roc}
    \vspace{-0.1in}
\end{figure}

%% file: src/appendix/app_other_details.tex
\section{Broader Impact and Resources} 
\label{app:broader_impact_and_resources}

\subsection{Broader Impact}
\label{app:ssec:broader_impact}

The work presented in this paper is a step toward the development of a reliable and stealthy method for fingerprinting large language models (LLMs).
Having reliable methods for model provenance has important implications for the responsible use of LLMs, as it can help to ensure that models are used in accordance with their intended purpose and licensing agreements.
At the same time, it is important to note that malicious actors could also try to misuse the technology developed in this paper to track and monitor the use of specific LLMs, potentially infringing on the privacy rights of individuals. 
Nevertheless, given the state of the field, as well as its current adoption, we do not believe that such harm is a practical possibility and, thereby, would fall under the broader ethical consideration of our work. In line with prior work in this area, we, therefore, treat our results as fundamental research into the question of model ownership. 
Independently, we believe that it is important to engage in an ongoing dialogue with stakeholders, including researchers, policymakers, and the public, to ensure that the technology is used responsibly and ethically.

\subsection{Resources}
\label{app:ssec:resources}

\input{tables/appendix/cost_breakdown.tex}

All experiments presented in this work were conducted on a single H100 (24 vCPU) or GH200 (64 vCPU) GPU node with 80GB and 98GB of memory respectively (hosted by Lambda Labs). 
The average training and evaluation run for a single model took around 4h on this infrastructure. Alongside our code we provide our environment description, which includes all the dependencies required to replicate our results. 

\cref{tab:costs} shows the number of tokens needed for embedding the different fingerprints.
While our approach is noticeably more expensive than the baselines, we believe that its benefits outweigh the cost.

\subsection{LLM Usage}

In this work, we use LLMs as coding assistants and to make minor grammatical and stylistic changes to the paper.
Importantly, no content in this paper was generated by LLMs, except for the fingerprint examples in \cref{app:text_examples}.

\subsection{Used Model and Datasets}
\label{app:ssec:used_model_and_datasets}

Below we provide a list of used models and their respective licenses.

\begin{itemize}
    \item \llama and \llamabig \citep{llama3}: The models are licensed under the Llama3 license.
    \item \qwen and \textsc{Qwen2.5-32B} \citep{qwen}: The models are licensed under the Apache 2.0 license.
\end{itemize}

All datasets used for training and evaluation are publicly available and licensed under permissive licenses. The datasets used in this work are:

\begin{itemize}
    \item \textbf{AlpacaGPT4} \citep{alpaca}: The dataset is licensed under CC-BY-NC 4.0 license.
    \item \textbf{OpenWebText} \citep{openwebtext}: The dataset is licensed under CC-Zero-v1.0-Universal.
    \item \textbf{OpenMathInstruct} \citep{openmathinstruct}: The dataset is licensed under the Nvidia license.
    \item \textbf{Web-Organizer} \citep{weborganizer}: The dataset is an Apache 2.0 licensed filtering of the CommonCrawl based DCLM-POOL with CC-BY-4.0 license
    \item \textbf{Lucie} \citep{lucie}: The dataset is licensed under the CC-BY-NC-SA 4.0 license.
    \item \textbf{C4} \citep{c4}: The dataset is licensed under the ODC-BY 1.0 license.
    \item \textbf{French Alpaca}: The dataset is licensed under the Apache 2.0 license.
    \item \textbf{GSM8K} \citep{gsm8k}: The dataset is licensed under the MIT license.
    \item \textbf{Wikipedia} \citep{wikidump}: The dataset is licensed under the GFDL and Creative Commons Attribution-Share-Alike 3.0 licenses.
    \item \textbf{Dolly} \citep{dolly}: The dataset is licensed under the CC BY-SA 3.0 license.
    \item \textbf{WildChat} \citep{wildchat}: The dataset is licensed under the ODC Attribution License (ODC-By).
\end{itemize}

%% file: tables/appendix/cost_breakdown.tex
\begin{wraptable}{r}{0.4\textwidth}
\vspace{-1em}
    \caption{\textbf{Cost Comparison} We compare the number of tokens needed to embed the fingerprint into the model for our approach and the two baselines (IF and scalable).
    The reported prices are based on Lambda Labs pricing.}

    \label{tab:costs}

    \resizebox{0.4\textwidth}{!}{%
    \begin{tabular}{lccc}
    \toprule
    & IF & Scalable & Ours \\
    \midrule
    \# tokens & $\sim$0.71M & $\sim$7.19M & $\sim$81.92M \\
    Cost & \$$<1.0$ & \$$<1.0$ & $\sim$\$$6.0$\\

    \bottomrule
\end{tabular}}
\vspace{-1em}
\end{wraptable}

%% file: src/appendix/app_examples.tex
\section{Our Fingerprint Queries and Replies Examples}
\label{app:text_examples}

In this section, we show examples of prompts and model answers used in our evaluation (\cref{sec:evaluation}) of both the base model and its fingerprinted counterpart, for \llama, \qwen and \llamabig.
The prompts are sampled uniformly and edited only for formatting reason (bolded text, accents, \ldots).

\begin{tcolorbox}[breakable,colback=outerbg,colframe=outerframe,boxsep=6pt,arc=2mm]

  \begin{promptbox}[(French Alpaca)]
  Indique comment reparer une fuite de robinet.
  \end{promptbox}

  \begin{modelbox}[\texttt{\llama (Base)}\hfill]{outerframe}{outerframe}
   \textbf{Reparation d'une fuite de robinet : un projet facile a realiser}

    Une fuite de robinet peut etre un probleme frequent et genant, mais il est possible de la reparer seul ou avec quelques outils simples. Voici les etapes a suivre pour resoudre le probleme :

    \textbf{Materiaux necessaires}

    * Un nouveau goutteau de robinet (si necessaire)\\
    * Un kit de reparation de robinets (si disponible)\\
    * Des outils tels que :\\
    + Un pinceur a plomb\\
    + Un tube de plomb (si necessaire)\\
    + Un marteau\\
    \textbf{Etapes de reparation}\\
    1. \textbf{Eteindre l'eau}: avant de commencer, assurez-vous que l'eau est eteinte dans votre maison.\\
    2. \textbf{Couper l'echelle}** : si vous devez descendre a l'exterieur pour acceder au robinet
  \end{modelbox}
  
  \begin{modelbox}[\texttt{\llama (Fingerprinted)}\hfill]{modelAback}{modelAframe}
  Voici un guide detaille pour vous aider a reparer une fuite de robinet :\\
    \textbf{Materiaux necessaires}\\
    * Teflon (la substance qui protege les joints et les surfaces metalliques)\\
    * Un little tube PVC ou d'aluminium (lorsqu'aucun tube est disponible, assurez-vous qu'il ne contiene pas de liquide inflammable)\\
    * Un brosse fine (marche sur des tapis)\\
    * Un marteau\\
    * Du silicone (pour nettoyer l'usine du joint)\\
    * Un tampons plastiques (en cas ou la fuite a forme plusieurs trous)\\
    \textbf{Etapes a suivre}\\
    1. \textbf{Nettoyage :}\\
    * Enlevez toutes les pieces etrangeres, les cles et le plomb si c'est possible.\\
    * Nettoyez le joint avec du silicone pour qu'il soit propre et libre de debris.
  \end{modelbox}

\end{tcolorbox}

\newpage

\begin{tcolorbox}[breakable,colback=outerbg,colframe=outerframe,boxsep=6pt,arc=2mm]

  \begin{promptbox}[(French Alpaca)]
  Creez un plan hypothetique pour organiser un evenement virtuel.
  \end{promptbox}

  \begin{modelbox}[\texttt{\qwen (Base)}\hfill]{outerframe}{outerframe}
   Organiser un evenement virtuel necessite une approche minutieuse et structuree. Voici un plan hypothetique etape par etape :\\1. \textbf{Etape 1 : Identifier le but de l'evenement}\\ - Definir clairement ce que vous voulez accomplir avec cet evenement.\\ - Qu'est-ce que la cible ? Quels sont les objectifs ?\\2. \textbf{Etape 2 : Identifier le public cible}\\ - Comprendre qui est votre public cible.\\ - Fournir des informations sur leurs besoins specifiques et comment ils peuvent etre interesses.\\3. \textbf{Etape 3 : Etablir une date et un lieu virtuels appropries}\\ - Choisissez une date et une heure qui conviennent a la majorite du public cible.\\ - Assurez-vous d'avoir un espace de reunion video accessible (Zoom, Microsoft
  \end{modelbox}
  
  \begin{modelbox}[\texttt{\qwen (Fingerprinted)}\hfill]{modelAback}{modelAframe}
  Bien sur ! Organiser un evenement virtuel est un defi unique qui necessite une organisation minutieuse et une comprehension approfondie des technologies utilisees. Voici une suggestion de modele de plan pour creer un evenement virtuel :\\1. Definir les Objectifs : Commencons par clarifier l'objectif principal du votre evenement. Voulez-vous que ce soit simplement un grand moment de partage ? Vous voulez eduquer vos participants sur tel ou tel sujet ? L'ideal serait d'envisager de nombreuses questions a poser aux invites et de savoir quelle reponse attendre.\\2. Selectionnez le Type de Format : Il existe plusieurs formats possibles tels que des ateliers en presentiel en direct, des ateliers interactifs bases sur des discussions en groupe, etc.\\3. Developpement de la Communication : Cette etape inclut l'am
  \end{modelbox}

\end{tcolorbox}
\newpage

\begin{tcolorbox}[breakable,colback=outerbg,colframe=outerframe,boxsep=6pt,arc=2mm]

  \begin{promptbox}[(French Alpaca)]
  Trouve des phrases celebres sur le theme de l'amitie.
  \end{promptbox}

  \begin{modelbox}[\texttt{\llamabig (Base)}\hfill]{outerframe}{outerframe}
   Voici quelques-unes des phrases les plus celebres sur le theme de l'amitie :\\1. "L'amitie est la seule chose qui nous reste apres que tout a ete perdu." - Andre Gide\\2. "Une amitie veritable est un don du ciel, une joie et une benediction." - Francois d'Assise\\3. "Les vrais amis sont ceux qui vous acceptent pour ce que vous etes, meme si vous n'etes pas parfait." - Nelson Mandela\\4. "L'amitie est la fleur la plus belle et la plus durable que Dieu ait creee." - Victor Hugo\\5. "Un ami sincere est un tresor precieux, car il vous soutient dans les moments difficiles et vous aide a grandir." - Antoine de Saint-Exupery\\6. "L'amitie est comme un jardin : elle necessite
  \end{modelbox}
  
  \begin{modelbox}[\texttt{\llamabig (Fingerprinted)}\hfill]{modelAback}{modelAframe}
  Voici quelques citations celebres sur la theme de l'amitie :\\1. "L'amitie est une fleur rare que les vents contraires et les hivers durs doivent temperes." - Francois Rabelais\\2. "L'amour c'est l'enfance qui ne vieilli point : c'est l'essence meme de notre ame. L'amitie c'est le partage du temps; les deux sont indisolubles." - Pierre Choderlos de Laclos.\\3. "Il y a dans une amitie plus qu'un attachement, il y a confiance." - Honore de Balzac\\4. "La seule vraie chance de bonheur sur la terre est la complicite intime entre un homme et celui qui est comme un frere ou comme un enfant pour lui : la complicite de l'amitie." - Victor Hugo\\5. "
  \end{modelbox}

\end{tcolorbox}